\documentclass[useAMS,usenatbib,usegraphicx]{mn2e}
\usepackage{graphicx,amssymb,amsmath}
\usepackage{color}
\newcommand{\teff}{$T_{\rm eff}$}
\newcommand{\logg}{$\log g$}
\newcommand{\logL}{$\log (L/L_\odot)$}

\newcommand{\oo}{HD\,25558}
\newcommand{\famias}{{\sc FAMIAS}}

\newcommand{\most}{{\it MOST}}
\newcommand{\corot}{{\it CoRoT}}
\newcommand{\kepler}{{\it Kepler}}
\newcommand{\gdor}{$\gamma$\,Dor}

\newcommand{\vsini}{$v \sin i$}

\newcommand{\pmodes}{$p$-modes}
\newcommand{\gmode}{$g$-mode}
\newcommand{\gmodes}{$g$-modes}

\newcommand{\cd}{d$^{-1}$}
\newcommand{\kms}{km\,s$^{-1}$}
\newcommand{\fmm}{\hbox{$.\!\!^{\rm m}$}}        

\newcommand{\pz}{\phantom{0}}                    

\title[Analysis of the SPB binary HD\,25558]
      {
          Extensive study of HD\,25558, a long-period double-lined binary with two SPB components
      }

\author[\'A. S\'odor et al.]
       {\'A. S\'odor$^{1,2}$, 
       P. De Cat$^{1}$, D.J. Wright$^{1,3}$, C. Neiner$^{4}$, M. Briquet$^{5}$, P. Lampens$^{1}$,
\newauthor{R.J. Dukes$^{6}$, G.W. Henry$^{7}$, M.H. Williamson$^{7}$, E. Brunsden$^{8}$, K.R. Pollard$^{8}$, }
\newauthor{P.L. Cottrell$^{8}$, F. Maisonneuve$^{8}$, P.M. Kilmartin$^{8}$,J. Matthews$^{9}$, T. Kallinger$^{10}$, }
\newauthor{P.G. Beck$^{11}$, E. Kambe$^{12}$, C.A. Engelbrecht$^{13}$, R.J. Czanik$^{14}$, S. Yang$^{15}$, }
\newauthor{O. Hashimoto$^{16}$, S. Honda$^{16,17}$, J.N. Fu$^{18}$, B. Castanheira$^{19}$, H. Lehmann$^{20}$, }
\newauthor{Zs. Bogn\'ar$^2$, N. Behara$^{21}$, S. Scaringi$^{11}$, H. Van Winckel$^{11}$, J. Menu$^{11}$, }
\newauthor{A. Lobel$^{1}$, P. Mathias$^{22}$, S. Saesen$^{23,11}$, M. Vu{\v{c}}kovi\'c$^{24,25,11}$, }
\newauthor{and the MiMeS collaboration}
\\
\\
        $^{1}$Royal Observatory of Belgium, Ringlaan 3, B-1180 Brussel, Belgium; e-mail: adam.sodor@oma.be\\
        $^{2}$Konkoly Observatory, Research Centre for Astronomy and Earth Sciences, Hungarian Academy of Sciences, Budapest, Hungary\\
        $^{3}$Department of Astrophysics and Optics, School of Physics, University of New South Wales, Sydney 2052, Australia\\
        $^{4}$LESIA, Observatoire de Paris, CNRS UMR 8109, UPMC, Universit\'e Paris Diderot, 5 place Jules Janssen 92190 Meudon, France\\
        $^{5}$Institut d'Astrophysique et de G\'eophysique, Universit\'e de Li\`ege, All\'ee du 6 Ao\^ut 17, B\^at B5c, 4000 Li\`ege, Belgium\\
        $^{6}$Department of Physics and Astronomy, The College of Charleston, Charleston, SC 29424, USA\\
        $^{7}$Center of Excellence in Information Systems, Tennessee State University, \\
        \ \ 3500 John A. Merritt Blvd., Box 9501, Nashville, TN 37209, USA\\
        $^{8}$Department of Physics and Astronomy, University of Canterbury, Private Bag 4800, Christchurch, New Zealand\\
        $^{9}$Department of Physics and Astronomy, University of British Columbia, 6224 Agricultural Road, Vancouver, BC V6T 1Z1, Canada\\
        $^{10}$Institute for Astronomy (IfA), University of Vienna, T\"urkenschanzstrasse 17, 1180 Vienna, Austria\\
        $^{11}$Instituut voor Sterrenkunde, K. U. Leuven, Celestijnenlaan 200D, B-3001 Leuven, Belgium\\
        $^{12}$Okayama Astrophysical Observatory, National Astronomical Observatory, Kamogata, Okayama 719-0232, Japan\\
        $^{13}$Department of Physics, University of Johannesburg, P.O. Box 524, Auckland Park, Johannesburg 2006, South Africa\\
        $^{14}$Department of Physics, Private Bag X6001, Potchefstroom Campus, North-West University, Potchefstroom 2520, South Africa\\
        $^{15}$Department of Physics and Astronomy, University of Victoria, Victoria, BC V8W 3P6, Canada\\
        $^{16}$Gunma Astronomical Observatory, Takayama-mura, Agatsuma, Gunma 377-0702, Japan\\
        $^{17}$Nishi-Harima Astronomical Observatory, Center for Astronomy, University of Hyogo,\\
        \ \ \ 407-2, Nishigaichi, Sayo-cho, Sayo, Hyogo 679-5313, Japan\\
        $^{18}$Department of Astronomy, Beijing Normal University, 19 Avenue Xinjiekouwai, Beijing 100875, China\\
        $^{19}$Department of Astronomy, The University of Texas at Austin, Austin, TX 78712, USA\\
        $^{20}$Th\"uringer Landessternwarte Tautenburg, 07778 Tautenburg, Germany\\
        $^{21}$Institut d'Astronomie et d'Astrophysique, Universit\'e Libre de Bruxelles,\\
        \ \ \ CP. 226, Boulevard du Triomphe, 1050, Bruxelles, Belgium\\
        $^{22}$Universit\'e de Toulouse; UPS-OMP; IRAP;  F-65000 Tarbes, France\\
        \ \ \ and CNRS; IRAP; 57, Avenue d'Azereix, BP 826, F-65008 Tarbes, France\\
        $^{23}$Observatoire de Gen\`eve, Universit\'e de Gen\`eve, Chemin des Maillettes 51, 1290 Sauverny, Switzerland\\
        $^{24}$European Southern Observatory, Vitacura, Santiago, Chile\\
        $^{25}$Astronomical Observatory, PO Box 74, 11060 Belgrade, Serbia\\
        \vspace{6cm}
       }

\begin{document}
\date{Accepted 2013 ..... Received 2013 ...}

\pagerange{\pageref{firstpage}--\pageref{lastpage}} \pubyear{2013}

\maketitle

\label{firstpage}

\begin{abstract}
We carried out an extensive observational study of the Slowly Pulsating B (SPB) star, \oo. The $\approx$\,2000 spectra obtained at different observatories, the ground-based and \most\/ satellite light curves revealed that this object is a double-lined spectroscopic binary with an orbital period of about 9 years. The observations do not allow the inference of an orbital solution. We determined the physical parameters of the components, and found that both lie within the SPB instability strip. Accordingly, both show line-profile variations due to stellar pulsations. Eleven independent frequencies were identified in the data. All the frequencies were attributed to one of the two components based on Pixel-by-pixel variability analysis of the line profiles. Spectroscopic and photometric mode identification was also performed for the frequencies of both stars. These results suggest that the inclination and rotation of the two components are rather different. The primary is a slow rotator with $\approx$\,6\,d period, seen at $\approx60\degr$ inclination, while the secondary rotates fast with $\approx$\,1.2\,d period, and is seen at $\approx20\degr$ inclination. Spectropolarimetric measurements revealed that the secondary component has a magnetic field with at least a few hundred Gauss strength, while no magnetic field can be detected in the primary.
\end{abstract}


\section{Introduction}
\label{sect:intro}

How do stars evolve? To answer this key question of astrophysics, we need to know the physical processes that rule their interiors. Stellar pulsations provide a unique way of understanding the internal structure of stars through characterization of excited modes revealed in photometric brightness and spectroscopic line-profile variations (LPVs). By matching the observed and theoretically predicted frequency spectrum, severe constraints can be obtained on, for example, the mass, the internal rotation law, the metallicity and the convection. Stellar pulsations are found across the whole H--R diagram. To get a global overview of stellar evolution, it is of utmost importance to perform in-depth studies for a wide variety of pulsating stars.

The slowly pulsating B (SPB) stars are a class of mid- to late-main-sequence B stars pulsating in high-radial-order, low-degree gravity modes (\gmodes; restoring force is buoyancy) with observed periods between 0.3 and 3 days \citep{Waelkens1991A&A...246..453W}. The amplitudes of their variations in photometry and radial velocity are typically of the order of a few millimagnitudes and a few \kms, respectively \citep{DeCat2002ASPC..259..196D}. The pulsations of SPB stars are driven by the opacity mechanism acting on the iron opacity bump around 200\,000\,K \citep{Dziembowski1993MNRAS.265..588D,Gautschy1993MNRAS.262..213G}. Their \gmodes\ probe the deepest layers of the star, which makes them very interesting from an asteroseismic point of view \citep{DeCat2007CoAst.150..167D}. Since most of the \gmode\ pulsators are multi-periodic, the observed variations have long beat-periods and are generally rather complex. Hence, large observational efforts are required for in-depth asteroseismic studies.

In-depth asteroseismic analyses are still rare because two conditions have to be satisfied simultaneously: a sufficient number of pulsation modes should be observed {\it and} they have to be well identified, which means that the horizontal degree, $\ell$, and the azimuthal number, $m$, of the spherical harmonics describing the pulsation modes should be determined. High-S/N, high-resolution spectroscopic observations of LPVs allow a determination of both $\ell$ and $m$ of the observed modes and put constraints on the inclination, $i$, and rotational parameters. Moreover, compared to photometry, modes with a higher-degree $\ell$ and/or a lower pulsation amplitude become detectable. This encouraged us in 2008 to start organizing dedicated spectroscopic follow-up campaigns for a sample of carefully-chosen main-sequence \gmode\ pulsators with a large spread in the projected rotational velocity ($v \sin i$), because 
we aim to investigate whether there exists any connection
between the $\ell$ and $m$ values of the excited modes and the $v \sin i$ of the star. The detection of such a relationship may allow theoreticians to revise their pulsation theories, which could drastically simplify the asteroseismic process of matching theoretical pulsation spectra to those observed, making successful as\-te\-ro\-seis\-mo\-lo\-gy achievable with a less detailed knowledge of a star's pulsation modes. The organization of the dedicated spectroscopic follow-up campaigns has been successful \citep{DeCat2009AIPC.1170..480D}. Each star was observed at least for one season.

However, only the ultra-precise and continuous photometric observations of space missions like \most, \corot\ and \kepler\ enable the detection of a huge number of low-amplitude frequencies, free from the well known 1~\cd\ aliasing problems encountered with single-site ground-based observations. For a successful asteroseismic study, it is crucial that the correct frequency values, accompanied with the identification of the corresponding pulsation modes, are provided to theoreticians for asteroseismic modelling. Moreover, preliminary results based on \kepler\ time series seem to suggest that the majority of the SPB and \gdor\ stars exhibit simultaneously excited \pmodes\ that probe layers closer to the surface \citep{Balona2011MNRAS.413.2403B,Grigahcene2010ApJ...713L.192G,Uytterhoeven2011A&A...534A.125U}. The combination of satellite photometry with multi-site ground-based spectroscopy is therefore the key to a successful asteroseismic investigation \citep{Handler2009ApJ...698L..56H}. Additional ground-based multicolour photometry allows an independent determination of $\ell$ for each pulsation frequency \citep{Dupret2003A&A...398..677D}

For magnetic stars, spectropolarimetry allows to study magnetic-field variations for a determination of the rotation period and the magnetic geometry, and hence the inclination of the rotation and magnetic axes, which could significantly narrow the free parameter space of the mode identification. It is also important to know if the star is magnetic for the seismic modelling and interpretation. Moreover, the individual spectra can be inserted in the spectroscopic data-sets used for LPV analysis.

The SPB star \oo\ (HIP\,18957, V1133\,Tau) was considered as the ideal target for an intense multi-year, multi-technique, multi-site and space campaign for several reasons. It is a bright (V\,= 5.3\,mag), and easily observable object from both hemispheres ($\alpha_{2000} = 04^{\rm h}03^{\rm m}44.\!\!^{\rm s}61$, $\delta_{2000} = +05{\degr}26'08.\!\!^{\prime\prime}2$). 
It shows a promising pattern of LPVs \citep{Mathias2001AA...379..905M}. Known to be a slow rotator (\vsini\,$\approx$\,22\,\kms; \citealt{Mathias2001AA...379..905M}), we can expect to avoid too many complications in the analysis induced by the effects of rotation.
Knowledge of the internal structure and evolution of such a massive star is of great importance for astrophysics because it forms the CNO elements.

\oo\ was discovered to be a SPB star by \cite{waelkens98}. Variability studies pointed out that this star has one dominant frequency of \hbox{0.653\,\cd}, but marginal detection of several further frequencies suggested multiperiodicity \citep{waelkens98,Mathias2001AA...379..905M,decat07,Dukes2009AIPC.1170..379D}. \citet{Hubrig2009AN....330..317H} published the detection of a longitudinal magnetic field of $\sim$\,100~Gauss in \oo. However, this result was later put in questions by \citet{Bagnulo2012A&A...538A.129B}. \oo\ was not known to be a multiple system before our study.

\section{Observations and data preparation}
\label{sect:data}

The observations in the framework of our project started in 2008. In this paper, we analyse the spectroscopic and photometric observations of \oo\ obtained up to the 2012 (spectroscopy) and 2013 (photometry) observing seasons. Since the observing season of \oo\ extends from the second half of a calendar year to the first part of the next year, we refer to the observing seasons with the calendar year in which they begin, all along this paper. We also collected and analysed all the previous photometric and spectroscopic observations on \oo\ we were aware of. The data obtained by our project on \oo\ are available upon request from the authors.

\begin{table*}
  \centering
  \caption{Log of the spectroscopic and photometric observations of \oo\ analysed in this paper, including literature data.\label{tbl:obslog}}
  \begin{tabular}{lllcccr}
    \hline
    \multicolumn{1}{c}{Observatory} & \multicolumn{1}{c}{Telescope} & \multicolumn{1}{c}{Instrument} & Wavelength                      & From                 & To                  & obs. \\
                                     &                               &                                & \multicolumn{1}{c}{range (nm)} & \multicolumn{2}{c}{(JD\,$-$\,2\,450\,000)} &  \#\\
    \hline
     \multicolumn{7}{c}{\it Spectroscopic observations}\\
    Observatoire Pic du Midi, France& 2.0\,m Bernard Lyot           & NARVAL                         & 380--885                        & 5513                 & 5550                & $^a$\,19 \\
    OHP, France                     & 1.5\,m                        & AURELIE                        & 412--413                        & 0852                 & 1164                & $^b$\,22 \\
    TLS, Tautenburg, Germany        & 2.0\,m Alfred Jensch          & Coud\'e echelle                & 470--735                        & 4718                 & 4788                & 70  \\
    SAAO, South Africa              & 1.9\,m                        & GIRAFFE                        & 440--655                        & 5518                 & 5531                & 166 \\
    Xinglong Observatory, China     & 2.2\,m                        & echelle                        & 560--935                        & 4750                 & 5519                & 102 \\
    OAO, Okayama, Japan             & 1.9\,m                        & HIDES                          & 395--770                        & 4752                 & 4844                & 58  \\
    GAO, Gunma, Japan               & 1.5\,m                        & GAOES                          & 480--665                        & 5442                 & 5522                & 76  \\
    MJUO, Tekapo, New Zealand       & 1.0\,m McLellan               & HERCULES                       & 380--800                        & 5501                 & 5529                & 425 \\
    CFHT, HI, USA                   & 3.6\,m CFHT                   & ESPaDOnS                       & 380--885                        & 5401                 & 5527                & $^a$\,12 \\
    DAO, Victoria, BC, Canada       & 1.2\,m                        & McKellar                       & 630--640                        & 4716                 & 4898                & 14  \\
    --- '' ---                      & --- '' ---                    & --- '' ---                     & 445--460                        & 5507                 & 5513                & 18  \\
    Fairborn Observatory, AZ, USA   & 2.0\,m AST (T13)              & fiber-fed echelle              & 495--695                        & 5486                 & 5644                & 572 \\ 
    McDonald Observatory, TX, USA   & 2.1\,m Otto Struve            & Sandiford (SES)                & 440--495                        & 5517                 & 5531                & 321 \\
    ESO, La Silla, Chile            & 1.2\,m Euler                  & CORALIE                        & 390--680                        & 3951                 & 4082                & $^b$\,11  \\
    ORM, La Palma, Spain            & 1.2\,m Mercator               & HERMES                         & 380--900                        & 5425                 & 6337                & 144 \\
\hline
    \multicolumn{7}{c}{\it Photometric observations}\\
    \most\/ satellite               & 15\,cm                        & CCD                            & wide band                       & 5502                 & 5523                & 71750 \\    
    SAAO, Sutherland, South Africa  & 50\,cm                        & PMT                            & Johnson $V$                     & 5521                 & 5531                & 87  \\
    Fairborn Obs., AZ, USA          & 75\,cm APT (T5)               & single-channel PMT             & Str\"omgren                     & 3031                 & 5638                & $^{b,c}\approx$\,2200 \\
    --- '' ---                      & 40\,cm APT (T3)               & single-channel PMT             &  Johnson $BV$                   & 6554                 & 6634          & 68 \\
    \hline
    \multicolumn{7}{l}{$^a$\,Spectropolarimetric observations, each consists of four sub-exposures.}\\
    \multicolumn{7}{l}{$^b$\,Data (partially) available already before the start of our dedicated multi-site campaign on \oo\ in 2008.}\\
    \multicolumn{7}{l}{$^c$\,Per band: $uvby$.}\\
  \end{tabular}
\end{table*}

\begin{figure*}
 \begin{center}
  \includegraphics[width=173mm]{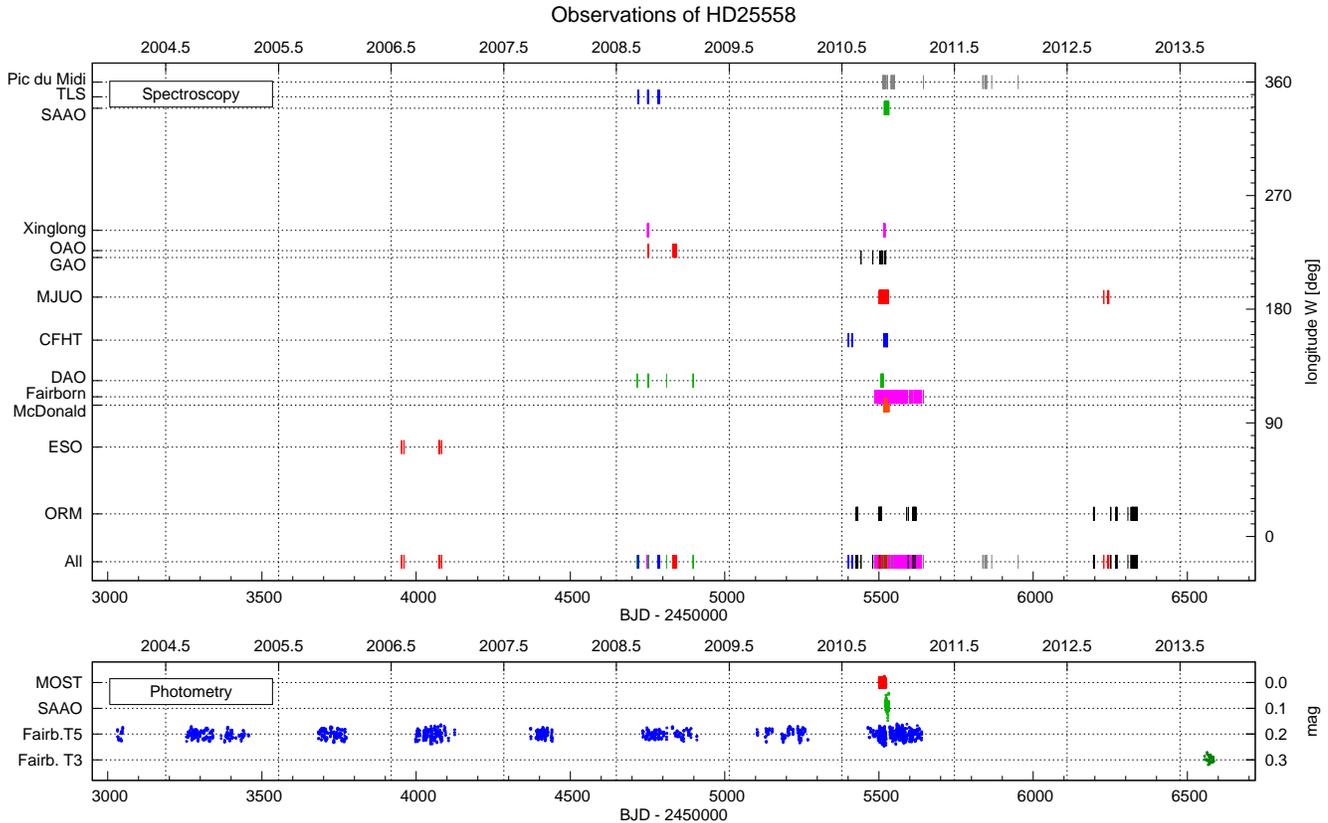}
 \end{center}
 \caption{Spectroscopic and photometric observations of \oo. The times and geographic longitudes of the spectroscopic observations are indicated in the upper panel. The lower panel delineates the light curves to demonstrate the time-distribution of these data. The middle of the calendar years (separating the observing seasons) are indicated on the top axes. Note that earlier literature data, obtained before 2004.0, are omitted from the plot.\label{fig:obslog}}
\end{figure*}

\subsection{Spectroscopy}
\label{sect:spect}

The time and geographical distribution of the spectroscopic observations are summarized in Table~\ref{tbl:obslog}, and are plotted in Fig.~\ref{fig:obslog}. The following abbreviations of the observatories are used in Table~\ref{tbl:obslog} and all along this paper:
Canada--France--Hawaii Telescope -- CFHT;
Dominion Astronomical Observatory -- DAO;
European Southern Observatory -- ESO;
Gunma Astronomical Observatory -- GAO;
Mount John University Observatory -- MJUO;
Observatoire Haut Provance -- OHP;
Observatorio del Roque de los Muchachos -- ORM;
Okayama Astrophysical Observatory -- OAO;
South African Astronomical Observatory -- SAAO;
T\'el\'escope Bernard Lyot -- TBL;
Th\"uringer Landessternwarte - TLS. 
Additionally to the observations obtained specifically for our project, we used earlier observations obtained with the AURELIE spectrograph (OHP, France -- \citealt{Mathias2001AA...379..905M}) and with the CORALIE spectrograph (ESO, Chile). Altogether, we have high-resolution spectroscopic data obtained with 14 different instruments in 6 observing seasons. The season-by-season distribution of the spectroscopic data is the following: 1998 -- 22; 2006 -- 11; 2008 -- 193; 2010 -- 1737; 2011 -- 9; 2012 -- 58 observations (counting the four spectropolarimetric sub-exposures as one observation, see end of Sect.~\ref{sect:spectropolarimetry} below).

The basic reduction of the spectroscopic observations, including the wavelength calibration, was made at the observatories with their own pipelines, with several exceptions. The McDonald and Xinglong observations were reduced and wavelength calibrated, and the OAO observations were wavelength calibrated by \'A. S\'odor using standard {\sc iraf}\footnote{{\sc iraf} is distributed by the National Optical Astronomy Observatory, which is operated by the Association of Universities for Research in Astronomy, Inc., under cooperative agreement with the National Science Foundation.} routines. The order-by-order normalization, merging and barycentric velocity correction were also done by \'A. S\'odor in some further cases. Finally, we filtered out the cosmics and normalized all the merged spectra in a standard way.

The spectroscopic data sets are not well suited for LPV analysis of the individual lines, since there is no suitable line in overlapping wavelength regions of most of the data sets, because of the heterogeneity of the instruments. Therefore, we rely on the cross-correlated mean line profiles in our LPV analysis.

The mean line profiles were calculated using a scaled-delta-function cross-correlation routine, which is the mathematical equivalent of constructing the weighted average of the selected line profiles in radial-velocity (RV) space. The weights were proportional to the equivalent widths (EW) of the lines, determined empirically from the averaged spectrum of the 2010 HERMES data. We used all the available subsets of 30 carefully selected strong but not heavily blended metallic lines for each spectrum. These spectral lines are listed in Table~\ref{tbl:linelist}. The blending was checked using the line-list output of a synthetic spectrum generated by {\sc Synspec v49}\footnote{http://nova.astro.umd.edu/Synspec49/synspec.html} \citep[][and references therein]{Hubeny1995ApJ...439..875H}, with the following parameters: $\mathrm{[Fe/H]=-0.3}$ \citep{Niemczura2003A&A...404..689N}, $T_\mathrm{eff} = 16\,600$\,K, \logg\,$ = 4.22$ (see Sect.~\ref{sect:relphyspar}), and using the atmosphere models of \cite{Castelli2003IAUS..210P.A20C}.

We tested the LPVs phase coherence between the lines used for cross-correlation by comparing LPV analysis results on subsets of lines of different ionization level of different species. The coherence was found to be satisfactory. We re-normalized the mean line profiles, and scaled the depths of the profiles of each instrument's data set to a common but arbitrary mean EW value to account for the differences arising from using different sets of lines for cross-correlation. The scaling factor was determined empirically from the EWs of the time-averaged mean line profiles of each instrument. Finally, we shifted the data sets to a common RV scale. The largest deviations from the mean RV zero point was detected for GAOES and Sandiford data, $-3.3$ and $+4.7$\,km\,s$^{-1}$, respectively, while in most of the cases, a smaller than 1\,km\,s$^{-1}$ shift was only necessary.

\begin{table}
  \centering
  \caption{Spectral lines used for cross-correlation and for relative physical parameter determination. Columns ``Wt.'' list the weights used for cross-correlation.\label{tbl:linelist}}
  \begin{tabular}{lrrlrr}
    \hline
       Elem. & \multicolumn{1}{c}{Wavel.}& \multicolumn{1}{c}{Wt.} & Elem. & \multicolumn{1}{c}{Wavel.} & \multicolumn{1}{c}{Wt.} \\
             & \multicolumn{1}{c}{(nm)}  &                         &       & \multicolumn{1}{c}{(nm)}\\
    \hline
    Si\,II   &   $^{\#\,}$385.6018 &   -- & Si\,II  &    $^{*\,}$505.6317 &   25 \\
    Si\,II   &   $^{\#\,}$386.2595 &   -- & Fe\,III &   $^{\#\,}$515.6111 &   -- \\
    Si\,II   &  $^{\#*\,}$412.8054 &  120 & Fe\,II  &  $^{\#*\,}$516.9033 &   50 \\
    Si\,II   &  $^{\#*\,}$413.0894 &  120 & Fe\,II  &   $^{\#\,}$526.0259 &   -- \\
    S\,II    &    $^{*\,}$416.2665 &   50 & Fe\,II  &    $^{*\,}$531.6615 &   25 \\ 
    S\,II    &    $^{*\,}$417.4002 &   45 & S\,II   &  $^{\#*\,}$532.0723 &   30 \\ 
    Fe\,II   &    $^{*\,}$423.3172 &   30 & S\,II   &    $^{*\,}$542.8655 &   35 \\ 
    C\,II    &    $^{*\,}$426.7261 &  175 & S\,II   &  $^{\#*\,}$543.2797 &   50 \\ 
    Fe\,III  &   $^{\#\,}$441.9596 &   -- & S\,II   &    $^{*\,}$545.3855 &   50 \\ 
    Mg\,II   &    $^{*\,}$448.1126 &  250 & Si\,II  &    $^{*\,}$546.6894 &   25 \\ 
    Al\,III  &   $^{\#\,}$451.2565 &   -- & S\,II   &  $^{\#*\,}$547.3614 &   30 \\ 
    Fe\,II   &  $^{\#*\,}$454.9474 &   30 & S\,II   &  $^{\#*\,}$560.6151 &   30 \\ 
    Si\,III  &    $^{*\,}$455.2622 &   60 & S\,II   &    $^{*\,}$563.9977 &   35 \\ 
    Si\,III  &  $^{\#*\,}$456.7840 &   40 & S\,II   &    $^{*\,}$564.0346 &   25 \\ 
    Si\,III  &  $^{\#*\,}$457.4757 &   25 & S\,II   &    $^{*\,}$564.7020 &   30 \\ 
    Fe\,II   &   $^{\#\,}$458.3837 &   -- & Ne\,I   &    $^{*\,}$614.3063 &   30 \\ 
    Fe\,II   &   $^{\#\,}$501.8440 &   -- & Si\,II  &    $^{*\,}$634.7109 &  140 \\ 
    Si\,II   &  $^{\#*\,}$504.1024 &   60 & Si\,II  &  $^{\#*\,}$637.1371 &   90 \\ 
    Si\,II   &   $^{*\,}$505.5984  &   75 & Ne\,I   &  $^{\#*\,}$640.2246 &   65 \\
    \hline
    \multicolumn{6}{l}{$\ ^*$: lines used for cross-correlation}\\
    \multicolumn{6}{l}{$^\#$: lines used for relative physical parameter}\\
    \multicolumn{6}{l}{\ \ \ determination (see Sect.~\ref{sect:relphyspar}.)}\\
  \end{tabular}
\end{table}

\subsubsection{Spectropolarimetry}
\label{sect:spectropolarimetry}

We obtained 31 spectropolarimetric measurements of \oo\ between July 2010 and January 2012: 12 measurements with ESPaDOnS at the CFHT in Hawaii and 19 with Narval at the TBL at the Pic du Midi observatory in France. The data have been collected in the frame of the Magnetism in Massive Stars (MiMeS) project \citep{Neiner2011sf2a.conf..509N}. Each measurement consists in 4 sub-exposures of 300 seconds for ESPaDOnS and 500 seconds for Narval taken in different configuration of the wave plates. The 4 sub-exposures are constructively combined to obtain the Stokes V spectrum in addition to the intensity spectrum. The sub-exposures are also destructively combined to produce a null profile to check for pollution by, for example, instrumental effects, variable observing conditions, or non-magnetic physical effects such as pulsations.

The usual bias, flat-field and ThAr calibrations have been obtained each night and applied to the data. The data reduction was performed using {\sc Libre-Esprit} \citep{donati1997}, a dedicated software available at TBL and CFHT. The intensity spectra were then normalized to the continuum level, and the same normalization was applied to the Stokes V and null spectra. 

We constructed a single averaged I and Stokes V profile for each measurement  applying the Least-Squares Deconvolution (LSD) technique \citep{donati1997}. These LSD I and Stokes V profiles have a much higher signal-to-noise (S/N) than individual lines, of about 1500 in I and between 7000 and 13\,000 in V on average per 2.6 km~s$^{-1}$ pixel.

For the LSD profiles, we computed two line masks by extracting line information from the VALD atomic database \citep{Piskunov1995A&AS..112..525P,Kupka1999A&AS..138..119K} for the VALD models the closest to the stellar parameters of each component of \oo, that is, [$T_{\rm eff}=17\,000$\, K, $\log g=4.0$] for the primary and [$T_{\rm eff}=16\,000$\,K, $\log g=4.5$] for the secondary (see Sect.~\ref{sect:physpar} for details). These masks originally included all lines with intrinsic line depths larger than 0.1. We then removed from the masks all lines that are not visible in the intensity spectra, H lines because of their Lorentzian broadening, those blended with H lines or interstellar lines, those with unknown Land\'e factors, lines in regions affected by absorption of telluric origin, as well as a few lines polluted by fringes. The depth of each line in the LSD mask was then adjusted so as to fit the observed depth. The final masks include 840 and 859 He and metallic lines, with averaged wavelength and Land\'e factors of [503.4 nm, 1.203] and [512.6 nm, 1.213], for the primary and secondary components, respectively.

We also used the average of the four sub-exposures of each spectropolarimetric observations together with the rest of the spectroscopic data, applying the same treatment, for the non-spectropolarimetric investigations.

\subsection{Photometry}
\label{sect:photometric_data}

Ground-based photometric observations were acquired in two observatories with three telescopes between the 2003 and 2013 seasons, using Str\"omgren $uvby$ and Johnson $BV$ filters. Space-based photometric observations were also performed by the \most\/ satellite. The log of the photometric observations can be found in the bottom part of Table~\ref{tbl:obslog}, and the time distribution of the light curve data are shown in the bottom panel of Fig.~\ref{fig:obslog}. Note that the first few seasons of the Str\"omgren data were already analysed by \cite{Dukes2009AIPC.1170..379D}. We also use previously analysed and published data from the {\it Hipparcos} satellite \citep{1997ESASP1200.....P}, and Geneva photometry from the Mercator Telescope \citep{decat07}.

\subsubsection{$MOST$ space photometry}

The \most\/ (Microvariability \& Oscillations of STars) satellite is a Canadian microsatellite equipped with a 15-cm telescope feeding a CCD photometer trough a custom broadband filter (350--700\,nm), capable of short-cadence, long-duration ultraprecise optical photometry of bright stars \citep{Walker2003PASP..115.1023W, Matthews2004Natur.430...51M}. \most\/ is in a Sun-synchronous polar orbit above the terminator at  820\,km altitude with an orbital period of about 101\,min. The data for \oo\ were obtained in the Direct Imaging mode, which is similar to conventional ground-based CCD photometry, and span a nearly continuous 21 day long interval in November 2010, with one major interruption of a few hours when the fine pointing of the satellite was lost. Individual exposures lasted 0.5\,s but were downloaded in ``stacks'' of 30 for the first about 7.5\,d of the observation, and stacks of 60 for the remaining 13.5\,d. Photometry was performed following the method of \cite{Rowe2006ApJ...646.1241R}, which combines classical aperture photometry and point-spread function fitting to the Direct Imaging subraster of the CCD. Images comprised by cosmic rays, image motion, or other problems were identified and removed. The final time series has 71\,750 data points. We applied further processing to remove the familiar periodic artefacts in the time series due to scattered Earthshine. First, we fitted a second-order polynomial to the measured background level of the magnitude of \oo\, and subtracted the fit from the magnitudes. The Fourier spectrum still had significant peaks at the orbital frequency of the satellite 
and its lower harmonics, as well as sidelobes arising from the amplitude modulation of the stray light by the Earth's rotation. For this reason, an additional correction was performed with the ``running averaged background'' method of \cite{Rucinski2004PASP..116.1093R}. For each day-long segment, the data were folded with the satellite's orbital period, boxcar-smoothed, and subtracted from the observed magnitudes. This suppressed the instrumental artefacts to only a small fraction of the intrinsic signal amplitudes. Further correction were applied during the Fourier analysis to remove the slight artificial brightness variations remaining in the data (see Sect.~\ref{sect:mostfourier}).

\subsubsection{Str\"omgren $uvby$ photometry from Fairborn Observatory}

The Str\"omgren differential photometric observations were obtained with the 75-cm T5 Four College Consortium Automatic Photoelectric Telescope (APT) located at Fairborn Observatory in Washington Camp, AZ, USA. Observations were made using the following procedure (standard for APT observations). The variable star being studied is compared with two reference stars designated comparison (comp) and check. These stars were respectively HD\,25490 (A1V, $V = 3.9$\,mag) and HD\,24817 (A2Vn, $V = 6.1$\,mag).  The four-colour sequence is similar to that for $UBV$ photometry as described by \citet{1984IAPPP..15...20B}. In this sequence, a single differential magnitude determination requires 44 individual 10-second measurements in the sequence: sky-comp-check-var-comp-var-comp-var-comp-check-sky. Each element in this sequence involves cycling through the four Str\"omgren filters. Additionally, one dark count was made after the four-filter sequence.

Since an absentee APT observer has relatively little information on the quality of a night, extra steps must be taken to eliminate measurements affected by cirrus clouds, etc. The analysis is begun by examining the magnitudes for quality after-the-fact \citep{duk91}. A common method, described in \citet{1986IAPPP..25...32H} and \citet{1988ApJS...67..439S}, is to discard observations whose comp minus check values differ by more than three standard deviations from their mean over the entire data set. One iterates this process until no more individual values qualify for rejection. The resulting standard deviation is taken as a measure of the precision of the photometry.

\subsubsection{Johnson $V$ photometry from SAAO}

The Johnson $V$ observations obtained with the PMT detector on the 0.5-m telescope at the Sutherland site of the SAAO were reduced by applying well-established dead-time corrections to the count rates, then using an E-region SAAO standard (E241 = HD\,24805, A3V, $V=6.896$\,mag) to fix the magnitude zero-point at the start of each night, and then using the same two comparison stars that were used for the Str\"omgren measurements to obtain differential photometry of \oo. The noise level in the final photometry of \oo\ was found to be substantially lower if only HD\,25490 was used as a comparison. Nightly variations in extinction were modelled by least-squares fitting of either a linear or a quadratic function (decided by visually inspecting the light curve of HD\,25490) to the HD\,25490 magnitudes over the night. All magnitudes were then corrected for the best-fitting extinction variation obtained on each night. Only two adjacent weeks of observing time were allocated on the 0.5-m telescope during the main \oo\ campaign in 2010, and useful data were only obtainable on seven nights in the 2-week period. \oo\ was setting by 2:20 am on these short southern summer nights, so the total yield of useful photometry for the 2-week period was only 21.5 hours.

Because of the unfavourable data distribution of the SAAO $V$ light curve, these data were used only for studying the \hbox{$O-C$} variations of the strongest periodicity. No other significant frequency can be detected in this data set.

\subsubsection{Johnson $BV$ photometry from Fairborn Observatory}

Between 19 September and 8 December in 2013, we acquired 68 observations with the T3 0.4-m (16-inch) APT, also located at Fairborn Observatory. T3 is one of eight automated telescopes operated by Tennessee State University at Fairborn for automated photometry, spectroscopy, and imaging \citep{Henry1995ASPC...79...44H,Henry1999PASP..111..845H,Eaton2003ASSL..288..189E,Eaton2007PASP..119..886E}. T3's precision photometer uses an EMI 9924B photomultiplier tube (PMT) to measure photon count rates successively through Johnson $B$ and $V$ filters. \oo\ was observed differentially with respect to a comparison star (HD~25621, $V=5.36$, $B-V=0.50$, F6 IV) and a check star (HD~25570, $V=5.45$, $B-V=0.37$, F2~V). The differential magnitudes were corrected for extinction and transformed to the Johnson $UBV$ system. To maximize the stability of the photometer, the PMT, voltage divider, pre-amplifier electronics, and photometric filters are all mounted within the temperature- and humidity-controlled body of the photometer. The precision of a single observation on a good night is usually in the range of $\sim$3--5\,mmag \citep[see, e.g.,][]{Henry1995ASPC...79...44H}, depending primarily on the brightness of the target and the airmass of the observation.

Similarly to the Johnson $V$ observations from SAAO, these small $B$ and $V$ data sets were used only to update the $O-C$ diagram of Fig.~\ref{fig:o-c} (see Sect.~\ref{sect:orbitphot}).

\section{Binarity}

\subsection{Orbital variations in spectroscopy}
\label{sect:orbitspect}
\begin{figure*}
 \begin{center}
  \includegraphics[width=180mm]{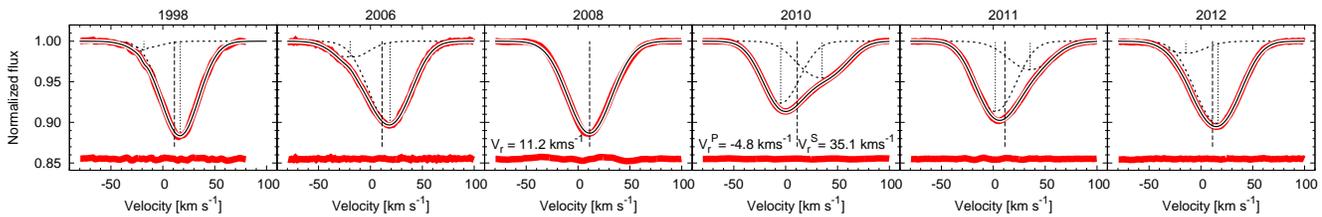}
 \end{center}
 \caption{Gaussian fits to the time-averaged cross-correlated line profiles of \oo\ in all the observing seasons. The profiles show the double-lined spectroscopic binary nature of the object. The observed mean line profiles are plotted with thick (red) lines. Thinner, black lines show the fitted functions of $2\times2$ co-axial Gaussians. The 2008 data were fitted with only one component, since the two lines almost completely overlap here. The components are also plotted separately with dashed lines. The residuals of the fits are shown in the bottom of each panel. Vertical dashed lines mark the centre-of-mass velocity of the system.\label{fig:doubleline}}
\end{figure*}

The time-averaged cross-correlated line profiles of the spectroscopic observing seasons, plotted in Fig.~\ref{fig:doubleline}, indicate that \oo\ is a double-lined binary (SB2). A weaker secondary component on the \hbox{left-,} \hbox{left-,} \hbox{right-,} right- and left-hand side of the primary component is apparent in the 1998, 2006, 2010, 2011 and 2012 season data, respectively. The mean profile of the 2008 season does not show an apparent double-line structure, here the profiles of the two components overlap almost completely. We fitted 2 co-axial Gaussians to each component's line profile to account for the slight deviations from the simple Gaussian profiles caused by, for example, rotational broadening. The residuals of the fits in the bottom of the panels of Fig.~\ref{fig:doubleline} show that these functions describe the profiles adequately.

There is no observable change in the positions of the lines of the two components within the observing seasons, indicating that the orbital period is on the order of several years. The available spectroscopic data are insufficient to determine the orbital period. In order to find a satisfactory orbital solution, we will continue to monitor this binary in the forthcoming seasons.

We adopt the fitted mean RV of the average line profile of 2008 as the centre-of-mass velocity of the binary star: $v_\mathrm{rad} = 11.2$\,\kms. This velocity is marked with vertical dashed line in each panel of Fig.~\ref{fig:doubleline}. The RVs of the components provided by the line-profile fits shown in Fig.~\ref{fig:doubleline} are reliable only for the most extended data of 2010, when the separation of two components' line profiles was the largest: $v_\mathrm{rad}^\mathrm{P} = -4.8$\,\kms, $v_\mathrm{rad}^\mathrm{S} = 35.1$\,\kms. Note that the superscripts P and S are used to denote quantities corresponding to the primary and secondary component, respectively, all along this paper. From these three RVs, we can roughly estimate the mass ratio of the system: $M^\mathrm{P} / M^\mathrm{S} \approx 1.5$.

In the 1998, 2006, 2011 and 2012 data, the fitted EW ratios of the components are quite different from those of the best observed 2010 season. The deviation in the 2006 and 2011 profiles can be explained by the scarce data of only 11 and 9 observations, respectively, so the LPVs are not averaged out quite well in these seasons. In the 1998 and 2012 average profiles, the RV separation of the two components is probably rather small, thus, the fit of the $2\times2$ Gaussians is not quite reliable. Also, the difference in the spectral type of the two components, and difference in the set of available lines used in the different instruments' data for cross-correlation, might explain some difference in the relative strength of the lines of the two components in these profiles.

\subsection{Orbital variations in photometry}
\label{sect:orbitphot}

Photometric observations of \oo\ are available on a longer time base and from more observing seasons than spectroscopic data. Previous studies revealed that the light variation of this object is dominated by one frequency of 0.652\,d$^{-1}$ \citep{waelkens98,decat07}, corresponding to a period of 1.532\,d. We refer to this frequency/period as the dominant frequency/period or dominant mode hereafter. Since the pulsation periods of SPB stars are known to be stable on the time scale of many years \citep{DeCat2002A&A...393..965D}, we assume that any phase change occurs mainly due to the light-time effect, therefore, the orbit can be studied via the $O-C$ diagram of the dominant period.

We constructed the $O-C$ diagram using all the available photometric data. We determined normal maximum timings from ``white-light'' brightness data of the multicolour Fairborn (Str\"omgren) and the previously published Mercator (Geneva) observations \citep{decat07} by calculating the average of the brightnesses for all times when data points were available from each band. We divided the light curves into observing seasons with the exception of the {\it Hipparcos} data \citep{1997ESASP1200.....P}, which is 2.1\,yr long but was considered as a single block, because of the uneven data distribution. We fitted the phase and amplitude of a fixed-period sine function, corresponding to the dominant pulsation period, to each light-curve segment. Normal maximum timings were calculated from the obtained phases.

\begin{figure}
 \begin{center}
  \includegraphics[width=90mm]{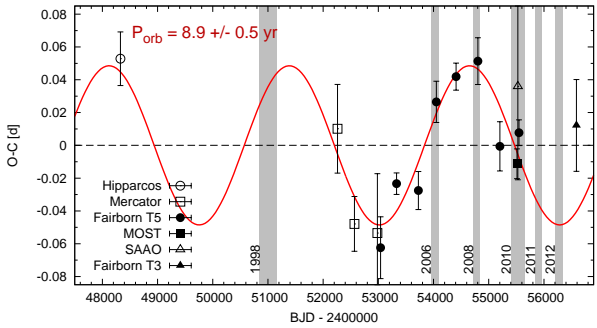}
 \end{center}
 \caption{$O-C$ diagram of the dominant period of \oo\ calculated for different photometric data sets. The error bars represent 1\,$\sigma$ uncertainties. The best-fitting sine curve corresponds to a 8.9-yr orbital period. The spectroscopic observing seasons are marked with vertical gray bands to help comparing the fitted sine curve with the average line profiles shown in Fig.~\ref{fig:doubleline}\label{fig:o-c}.}
\end{figure}

The $O-C$ diagram, shown in Fig.~\ref{fig:o-c}, was constructed using the following ephemeris:
$$\mathrm{BJD}_\mathrm{max} = T_0 + P_\mathrm{d} \cdot E,$$
$$\mathrm{where}\ T_0 = \mathrm{BJD}\ 2453001.1512\ \mathrm{and}\ P_\mathrm{d} = 1.53232423\,\mathrm{d.}$$

\noindent Here $T_0$ is an arbitrary light-maximum time of the dominant pulsation period, $P_\mathrm{d}$ is the mean period best describing the whole data set, and $E$ is the epoch number. The dashed line in Fig.~\ref{fig:o-c} represents a weighted linear fit to the plotted $O-C$ data. Our choice of $T_0$ and $P_\mathrm{d}$ ensures that this line runs horizontally at $O-C=0$. After setting these two parameters, we fitted a sine function to the $O-C$ data. The period of this sine curve is an estimation of the orbital period: $P_\mathrm{orb} = 8.9 \pm 0.5$\,yr.

The local slope of the $O-C$ curve, if caused by the light-time effect, corresponds to the instantaneous RV of the component that pulsates with the investigated period, relative to the centre of mass of the system \citep[see a more detailed analysis of the question by][]{2012MNRAS.422..738S}. A positive slope means that the light delay increases as the pulsating component moves away from us, while negative slope corresponds to a component approaching us. We marked the spectroscopic observing seasons with gray bands in Fig.~\ref{fig:o-c}. Comparing the slope of the $O-C$ curve in these intervals with the relative RVs of the components at different epochs, shown in Fig.~\ref{fig:doubleline}, we can deduce that the dominant light variation originates from the primary component of the binary.

A simple sine curve in the $O-C$ diagram corresponds to a circular orbit. The fitted sine curve runs through the 1\,$\sigma$ error bar of almost each $O-C$ data point, hinting to a nearly circular orbit. Nevertheless, the moderate number of data points does not permit the fitting of any higher-order curve, thus, we are unable to investigate the eccentricity in a quantitative way. We will also continue the photometric monitoring of \oo\ for finding an orbital solution.

Since the orbital phase variations, caused by the light-time effect, satisfactorily explain the variations in the $O-C$ diagram during the full 20-yr time span, our results support the long-term stability of the pulsation frequencies of SPB stars.

\section{Physical parameters of the binary components}
\label{sect:physpar}

\subsection{Average temperature, luminosity and \logg\ of the system}

Several earlier studies published atmospheric parameters of \oo. These were determined from multicolour Geneva photometry \citep{waelkens98, decat07, Hubrig2009AN....330..317H} and from spectroscopy \citep{Mathias2001AA...379..905M, Lefever2010A&A...515A..74L}. However, the binary nature of the system was not known at that time, thus, those parameters should be treated with caution. The published effective temperatures, $T_\mathrm{eff}$, fall between 16\,400 and 17\,500\,K, the logarithm of the luminosity in Solar units, $\log (L/L_\odot)$, between 2.76 and 2.81, and the logarithm of the surface gravity in cgs units, $\log g$, between 4.21 and 4.22.

According to line-profile fittings, the EW of the time-averaged mean line profile of the primary is about 35 per-cent larger than that of the secondary in the 2010 data (see the fitted curves in Fig.~\ref{fig:doubleline}). Visual inspection of time-averaged spectra of this season show that there are only little deviations from this mean EW ratio in the individual lines, suggesting that the primary is about 35 per-cent more luminous than the secondary, while the temperature difference between the components is quite low, probably less than 1000\,K. Therefore, \teff\ and \logg\ values obtained by photometry are acceptable approximations as the luminosity-weighted mean atmospheric parameters of the components.

Since earlier photometric studies assumed Solar metallic abundances, we re-determined the mean values of \teff\ and \logg\ using the published Geneva photometry \citep{decat07} and the metallicity value of $\mathrm{[Fe/H]}=-0.3$ \citep{Niemczura2003A&A...404..689N}. Mean magnitudes in the Geneva bands were determined by fitting the magnitude zero point of a single sinusoidal function of the dominant pulsation frequency. Using the calibration grid and interpolating software of \cite{Kunzli97}, we obtained $T_\mathrm{eff} = 16\,600\pm800$\,K and $\log g = 4.22\pm0.2$\,dex for the system. Note that here we adopted the more realistic error ranges of \cite{decat07}, instead of using the interpolation errors yielded by the software.

\begin{figure*}
 \begin{center}
  \includegraphics[width=175mm]{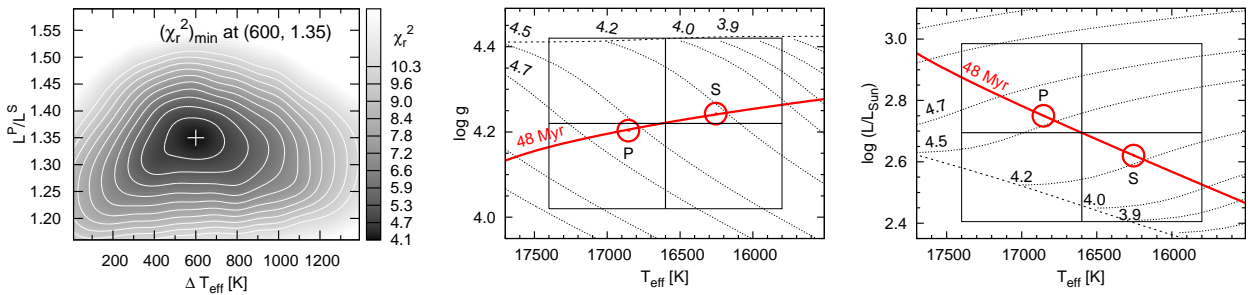}
 \end{center}
 \caption{Relative physical parameters of the two components of \oo. The left-hand panel shows the $\chi^2$ map for the relative luminosity and temperature difference of the components. The middle and right-hand panels show evolutionary tracks of CL\'ES (\citealt{Scuflaire2008Ap&SS.316...83S}; thin dotted lines with corresponding masses given in Solar units), the isochrone of 48\,Myr (thick gray/red line), the error boxes of the parameters derived by photometry, and the calculated positions of the primary (P) and secondary (S) components in this parameter space. Evolution along the tracks progresses from left to right, towards decreasing temperatures. The ZAMS is plotted with a dashed line.
 \label{fig:relphyspar}}
\end{figure*}

\subsection{Temperature difference, luminosity ratio, \logg\ difference and mass of the two components}
\label{sect:relphyspar}

We investigated the average of the 67 HERMES spectra \citep[for details of the instrument, see][]{Raskin2011A&A...526A..69R} observed in the 2010 season to derive the relative luminosity and temperature difference of the components. This average spectrum has the largest signal-to-noise ratio of all the spectra (at 500\,nm, S/N\,$\approx$\,2000 per wavelength bin corresponding to R\,$\approx$\,85\,000 resolution), and the number of observations are sufficient to average out the LPV. Only data from 2010 can be used for this kind of investigation, because the RV separation of the line profiles of the two components was sufficiently large only in this season to permit a comparative investigation.

We determined the EW ratios of the two components for 21 non-blended metallic lines by fitting the depths of $2\times2$ co-axial Gaussians functions to them. During these fits, we kept fixed the mean RVs, the width parameters and the relative depths of the two Gaussian components describing the profile of one stellar component, as determined by the fit to the time-averaged mean line profile from 2010 (plotted in Fig.~\ref{fig:doubleline}). In this way, only two depth parameters were fitted to each individual line, characterizing the EWs of the primary and the secondary. The spectral lines used for this investigation are listed in Table~\ref{tbl:linelist}. Note that the high S/N of the averaged HERMES spectrum from 2010 permitted the investigation of several weak and/or blue lines that were otherwise not used for the cross-correlation, because they would not improve the S/N of the resulted mean line profiles.


We assume that the chemical composition of the two components are identical, thus, EW differences in individual lines are caused only by luminosity, temperature and \logg\ differences. We also assume that the two components have the same age, and, as the orbit is quite wide and not eccentric, that the components have not affected each other's evolution. We calculated the isochrones crossing the (16\,600\,K, 4.22) point in the (\teff, \logg) plane, using stellar models computed with the evolutionary code CL\'ES (Code Li\'egeois d'\'Evolution Stellaire; \citealt{Scuflaire2008Ap&SS.316...83S}) with the input physics described in \cite{Briquet2011A&A...527A.112B}, assuming $X=0.7$ H abundance, $Z=0.01$ metallicity, using $\alpha = 1.8$ mixing-length parameter and three different overshooting parameters. These resulted in estimated ages of 40, 48 and 55\,Myr for the system for 0.0, 0.2 and 0.4 overshooting parameters, respectively. The uncertainty of the age is quite large, since the zero-age main sequence (ZAMS, independent of the overshooting parameter) is also within the error box (see the middle panel of Fig.~\ref{fig:relphyspar}). Note that the morphology of these three isochrones around the mean physical parameters of \oo\ are almost identical, therefore, we plotted only the 48-Myr isochrone corresponding to 0.2 overshooting parameter. We also plotted only the evolutionary tracks of this overshooting value in the middle and right-hand panels of Fig.~\ref{fig:relphyspar}. The two components are assumed to lay on this isochrone, surrounding the photometric mean physical parameters.

We computed synthetic spectra of $\mathrm{[Fe/H]}=-0.3$ between \teff\,=\,15\,800 and 17\,200\,K, with a step size of 100\,K, using {\sc Synspec} \citep{Hubeny1995ApJ...439..875H} with the atmosphere models of \cite{Castelli2003IAUS..210P.A20C} interpolated linearly between the original grid points. The \logg\ values for the model spectra were selected from a narrow range between 4.26 and 4.18, according to the obtained isochrone (see the middle panel of Fig.~\ref{fig:relphyspar}).

We selected pairs from these synthetic spectra with \teff\ differences ($\Delta T_\mathrm{eff}$) in the range of 0--1400\,K, using 100\,K steps, and scaled them according to different relative luminosities ($L^\mathrm{P}/L^{\mathrm S}$) in the range of 1.1--1.6, using a step size of 0.05. The pairs were always selected from this grid in such a way that their luminosity-weighted average temperature was as near to 16\,600\,K as possible. 

Theoretical EW ratios were then calculated from these models for the 21 spectral lines under investigation. We compared these theoretical values with the observed ones by calculating the reduced chi-square ($\chi^2_\mathrm{r}$) for each ($\Delta T_\mathrm{eff}$, $L^\mathrm{P}/L^{\mathrm S}$) pair to find the best-fitting parameters.

The results of these calculations are shown in Fig.~\ref{fig:relphyspar}. The $\chi^2_\mathrm{r}$ map in the left-hand panel shows that the temperature difference between the two components is small indeed, as expected. The best solution has a goodness of $\chi^2_\mathrm{r}=4.1$. The contours in this panel show 15 per-cent increments in $\chi^2_\mathrm{r}$ (see the scale in the grayscale-box), therefore, the innermost contour corresponds to 95 per-cent confidence level, that is, about 2\,$\sigma$ uncertainty. According to the best solution, the primary is warmer only by about $600\pm150$\,K, and is about $1.35\pm0.05$ times more luminous than the secondary component.

Among the 21 spectral lines we used for this investigation, there are lines with negative, positive and almost neutral EW\,--\,\teff\ dependence, thus, the determined $\Delta T_\mathrm{eff}$ and $L^\mathrm{P}/L^{\mathrm S}$ are practically uncorrelated, as the left-hand panel of Fig.~\ref{fig:relphyspar} demonstrates. We also note that the temperature difference determined this way is more accurate than the photometric measurement of the average temperature itself.

Evolutionary tracks and the isochrone of \oo\ are plotted together with the photometric mean parameters and their error ranges in the middle and right-hand panels of Fig.~\ref{fig:relphyspar}. The locations of the two components, taking into account their 600\,K temperature difference, the 1.35 luminosity ratio and the luminosity-averaged mean photometric values, are marked in these panels.

Considering the theoretical evolutionary tracks plotted in Fig.~\ref{fig:relphyspar}, the masses of the two components are \hbox{$M^\mathrm{P} \approx 4.6\,M_\odot$} and \hbox{$M^\mathrm{S} \approx 4.2\,M_\odot$}. Note that changing the overshooting parameter by $\pm0.2$ changes the derived masses by less than $\pm0.1\,M_\odot$. These masses yield a mass ratio of only $M^\mathrm{P}/M^\mathrm{S} \approx 1.1$. There is a discrepancy between this value and the mass ratio of $\sim$\,1.5 estimated tentatively from the RVs of the components' lines in Sect.~\ref{sect:orbitspect}. A shift of about $+3$\,km/s in the mean RV could resolve this discrepancy. Such a shift might originate from instrumental effects, and also the profiles of the two components might not completely overlap in the 2008 season, contrary to the assumption we made when determining the centre-of-mass velocity of the system.

Our best estimate of some of the physical parameters of the two components is:
\begin{gather*}
T_\mathrm{eff}^\mathrm{P} = 16\,850\pm{800}\,\mathrm{K}, \ \ \ \ T_\mathrm{eff}^\mathrm{S} = 16\,250\pm{1000}\,\mathrm{K},\\
\log g^\mathrm{P} = 4.2\pm{0.2}, \ \ \ \ \log g^\mathrm{S} = 4.25\pm{0.25},\\
\log (L^\mathrm{P}/L_\odot) = 2.75\pm{0.29}, \ \ \ \ \log (L^\mathrm{S}/L_\odot) = 2.62\pm{0.36}.
\end{gather*}

\noindent Note that the uncertainties for the primary are adopted from \cite{decat07}, while those for the secondary were increased by 25 per-cent to account for the larger uncertainties caused by the lower luminosity of this component.

Because of the low difference in \teff\ and \logg\ between the two components of \oo, both stars are located in the theoretical SPB instability region of the HRD (see, e.g., fig.~1c in \citealt{decat07}), consequently, both are expected to exhibit stellar pulsations.

\subsection{Spectropolarimetric measurement of the magnetic field}

\begin{figure}
\begin{center}
  \includegraphics[width=0.92\hsize]{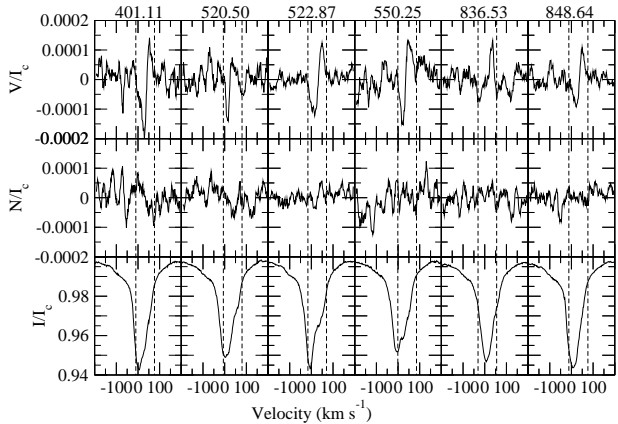}
\caption[]{Examples of LSD profiles of \oo\ showing a Stokes V signature of the presence of a magnetic field. Normalized Stokes V (top), null N (middle) and intensity I (bottom) profiles are shown for each measurement. Vertical dashed lines delimit the signature width. The times of mid-exposures (BJD\,$-$\,2\,455\,000) are indicated above the columns. Note that the first four observations were obtained in 2010, while the last two in 2011.
}
\label{LSDprof}
\end{center}
\end{figure}

Examples of LSD profiles computed with the mask optimized for the secondary component (see Sect.~\ref{sect:spectropolarimetry}) are shown in Fig.~\ref{LSDprof}.

The null profiles are noisy but mostly flat, which shows that the magnetic measurements have not been polluted by spurious polarization or pulsational line-profile changes between the sub-exposures.
Some of the Stokes V profiles, however, show signatures that indicate the presence of a magnetic field in \oo. These signatures seem to be centered on the secondary intensity profiles, while no signature can be detected for the primary component. Therefore, we conclude that most probably the secondary component of \oo\ is magnetic, while no field is detected in the primary with the achieved detection level.

Extraction of the precise longitudinal magnetic field value, $B_l$, and thus of the magnetic field strength and geometrical configuration would require disentangling of the intensity spectra. This has not been possible with our current knowledge on the orbit, therefore, we cannot determine the magnetic field parameters. Using the full (primary+secondary) intensity profile, however, and assuming an integration domain between $-10$ and 90\,\kms\ for the secondary component, we can determine a lower limit of the longitudinal field value. This value is a lower limit because the Stokes V profiles are normalised by a too strong intensity corresponding to the contribution of the primary and the secondary components rather than to the intensity of the magnetic star. We find that $B_l$ varies between $-54$ and 32\,G, with a typical error bar of 15\,G. Considering that these values are underestimates of the real longitudinal field, the maximum $|B_l|$ can be estimated to be of the order of $\sim100$\,G. Following \cite{Schwarzschild1950ApJ...112..222S}, the polar field strength of the secondary component of \oo\ can be estimated to be 3.16 times $|B_l|$, that is, of a few hundred Gauss


\section{Frequency analysis}

We looked for a mathematical description of the variations of different photometric and spectroscopic observables in the form of Fourier sums, applying discrete Fourier transformation and non-linear and linear least squares fitting methods utilising the {\sc LCfit} \citep{lcfit}, \famias\ \citep{famias} and {\sc MuFrAn} \citep{mufran} program packages.

A peak in the Fourier spectrum is accepted as intrinsic when its amplitude exceeds the usually accepted limit of 4.0\,$\sigma$ \citep{Breger1993A&A...271..482B}, where $\sigma$ is the average of the amplitude spectrum in a given vicinity of the peak in question. We also give the S/N value for the amplitude of each identified frequency component, where the noise is estimated as the $\sigma$ of the residual spectrum around the given frequency after prewhitening the data with all the significant frequencies identified.

We weighted each data point equally in the time series during the Fourier analysis of the photometric data. During the Fourier analysis of the spectroscopic time series, each data point was weighted with the empirically determined S/N of the corresponding spectrum.

\subsection{Photometry}

\subsubsection{\most\/ photometry}
\label{sect:mostfourier}

\begin{figure}
 \begin{center}
  \includegraphics[width=77mm]{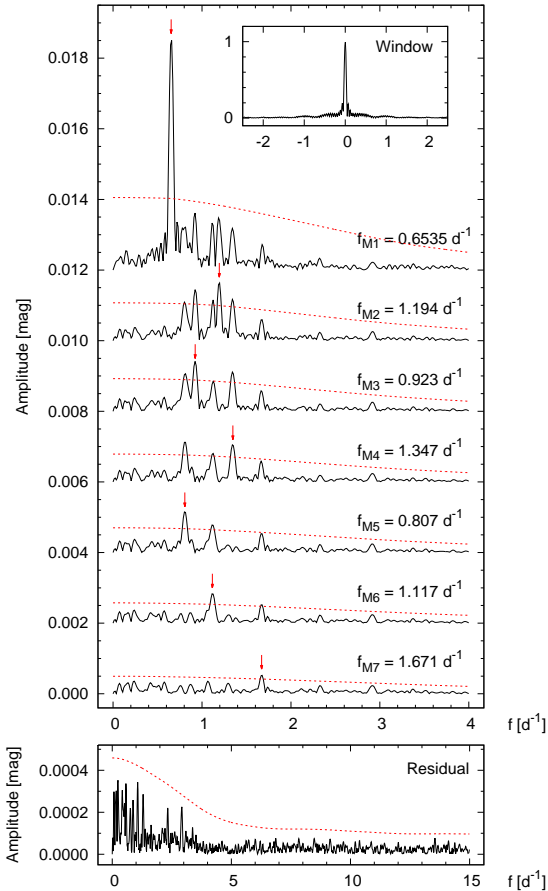}
 \end{center}
 \caption{Prewhitening steps of the \most\/ light curve of \oo. The top panel shows the prewhitening steps, while the residual spectrum is shown in the bottom panel for a larger frequency range. The spectral window function in the insert demonstrates that there are basically no alias peaks in the spectra. Dashed lines represent the 4.0\,$\sigma$ noise level for each step.\label{fig:mostfourier}}
\end{figure}

\begin{figure*}
 \begin{center}
  \includegraphics[width=177mm]{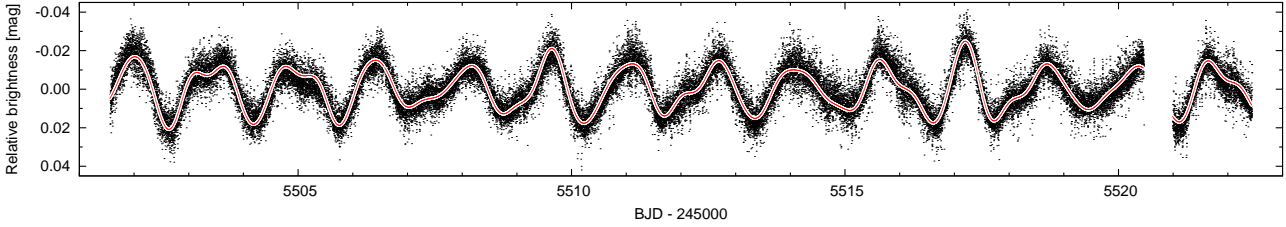}
 \end{center}
 \caption{The filtered $MOST$ light curve of \oo\ and the fitted 7-frequency solution.\label{fig:mostlcfit}}
\end{figure*}

As we already mentioned in the data description, the cadence time of the \most\/ observations was $\approx$\,16\,s before BJD\,2\,455\,509.615 and $\approx$\,32\,s afterwards. To balance the weights of the single data points, we binned the shorter-cadence data points by 2. After this step, the light curve contained 51\,602 data points.

Some systematic instrumental effects were not completely removed by the data reduction process described in Sect.~\ref{sect:photometric_data}. The most important of these are caused by the scattered light of the Moon, since the almost full Moon passed by $<20\degr$ from \oo\ during the observing run. Another important, not completely removed contamination factor is the scattered light reflected from the surface of Earth. This causes variations in the measured brightness of the target with the orbital frequency of the satellite ($f_\mathrm{orb} = 14.1994$\,d$^{-1}$). This light contamination is modulated by the synodic rotation frequency of the Earth ($f_\mathrm{E} = 1.0000$\,d$^{-1}$) due to the different-albedo surface features. Peaks at $f_\mathrm{E}$ and its harmonics appear directly in the Fourier spectrum. Furthermore, we have found that the light contamination is also modulated by the synodic orbital period of the Moon ($f_\mathrm{L} = 0.03386$\,d$^{-1}$).

All these effects add a complex artificial peak structure to the Fourier spectrum, because many high-order linear combinations of these frequencies emerge. We removed these signals from the light curve by a two-step iterative process. First, we determined the significant periodicities intrinsic to the star, and prewhitened the light curve with these variations. Then we fitted the residuals with the following independent and linearly dependent frequencies: $f_\mathrm{L}$, $n f_\mathrm{E}$ where $n = \{1, 2, ... 5\}$, $i f_\mathrm{orb} \pm j f_\mathrm{E} \pm k f_\mathrm{L}$ where $i = \{1, 2, ... 10\}$, $j = \{0, 1, 2, 3\}$ and $k = \{0, 1\}$, and $4 f_\mathrm{orb} \pm 2f_\mathrm{L}$. Note that these are the linear combinations of the mentioned artificial frequencies that we found by visual inspection up to the vicinity of the 10th harmonic of $f_\mathrm{orb}$. Next, we subtracted this 218-frequency solution from the original \most\/ data, resulting in the filtered light curve of \oo. Finally, we started over to identify the intrinsic frequencies of our object in the filtered data set.

The analysis of the filtered data revealed seven significant frequencies with $\mathrm{S/N} > 4.0$. These are listed in Table~\ref{tbl:mostfreq}. The prewhitening process is demonstrated in Fig.~\ref{fig:mostfourier}, and the light curve with the fitted solution is plotted in Fig.~\ref{fig:mostlcfit}. The seven-frequency fit of the data resulted in 5.66\,mmag rms.

The residual spectrum in the bottom panel of Fig.~\ref{fig:mostfourier} shows increased noise below about 3\,d$^{-1}$. A significant part of this noise is most probably the result of numerous low-amplitude signals in the data, intrinsic to the star, many of them are probably unresolved due to the limited length of the data set (21\,d).

It is important to note that the final number of identified significant frequencies depends strongly on the way we calculate the noise level. Furthermore, the final S/N estimation depends also on the number of identified significant frequencies, since the $\sigma$ of the residual spectrum is decreased by each further frequency prewhitened. The whole process is largely sensitive to the radius of the smoothing window, from which the spectral noise is estimated and the S/N of the next strongest peak is calculated after each prewhitening step. Due to the low resolution of the Fourier spectrum of the \most\/ data, a relatively large smoothing radius of $\pm2$\,\cd\ was adopted. Consequently, many peaks of real low-amplitude signal components might contribute to the noise. The listed seven frequencies are the result of a conservative selection criteria. These are detected with any reasonably large smoothing radius.

\begin{table}
  \centering
  \caption{Frequencies identified in the $MOST$ light curve of \oo. The standard errors of the fitted frequencies calculated by {\sc LCfit} \citep{lcfit} are given in parentheses in the unit of the last digit.\label{tbl:mostfreq}}
  \begin{tabular}{clrr}
    \hline
    ID & Frequency & Amplitude & S/N \\
       & \multicolumn{1}{c}{(d$^{-1}$)} & \multicolumn{1}{c}{(mmag)} \\
    \hline
    $f_{\mathrm{M}\,1}$ & 0.6535(2) & 13.1 & 59.1 \\
    $f_{\mathrm{M}\,2}$ & 1.194(1)  &  2.8 & 13.8 \\
    $f_{\mathrm{M}\,3}$ & 0.923(1)  &  2.9 & 13.6 \\
    $f_{\mathrm{M}\,4}$ & 1.347(1)  &  2.2 & 10.8 \\
    $f_{\mathrm{M}\,5}$ & 0.807(1)  &  2.5 & 11.3 \\
    $f_{\mathrm{M}\,6}$ & 1.117(2)  &  1.8 &  8.7 \\
    $f_{\mathrm{M}\,7}$ & 1.671(3)  &  1.1 &  5.7 \\
    \hline
  \end{tabular}
\end{table}

\begin{table*}
  \centering
  \caption{Frequencies identified in the filtered Fairborn light curves of \oo\ in four Str\"omgren bands, $uvby$, their uncertainties, the fitted amplitudes and the corresponding S/N values. Frequency uncertainties are calculated as the scatter of the values obtained for the four passbands. Uncertainties are given in parentheses in the unit of the last digit.\label{tbl:fairbornfreq}}
  \begin{tabular}{clrrrrrrrr}
    \hline
    ID & Frequency & Amplitude & S/N & Amplitude & S/N & Amplitude & S/N & Amplitude & S/N \\
       & \multicolumn{1}{c}{(d$^{-1}$)} & \multicolumn{1}{c}{(mmag)} & \multicolumn{1}{c}{(d$^{-1}$)} & \multicolumn{1}{c}{(mmag)} & 
         \multicolumn{1}{c}{(d$^{-1}$)} & \multicolumn{1}{c}{(mmag)} & \multicolumn{1}{c}{(d$^{-1}$)} & \multicolumn{1}{c}{(mmag)} \\
       & & \multicolumn{2}{c}{$u$} & \multicolumn{2}{c}{$v$} & \multicolumn{2}{c}{$b$} & \multicolumn{2}{c}{$y$} \\
    \hline
    $f_{\mathrm{Fb}\,1}$ & 0.652593(3) & 25.0(2) & 70.1 & 16.6(3) & 50.5 & 15.4(3) & 47.7 & 14.5(4) & 38.4 \\
    $f_{\mathrm{Fb}\,2}$ & 0.92277(2)  &  5.6(2) & 15.0 &  3.2(3) &  9.6 &  2.6(3) &  8.5 &  3.0(4) &  9.0 \\
    $f_{\mathrm{Fb}\,3}$ & 1.12909(1)  &  3.6(2) & 10.6 &  2.8(3) &  8.6 &  2.9(3) &  9.1 &  3.3(4) &  8.1 \\
    $f_{\mathrm{Fb}\,4}$ & 1.19184(5)  &  3.2(3) &  8.9 &  2.3(3) &  7.7 &  2.1(4) &  7.5 &  1.8(5) &  3.1 \\
    $f_{\mathrm{Fb}\,5}$ & 0.81106(8)  &  1.2(3) &  3.4 &  1.4(3) &  4.4 &  2.1(4) &  4.5 &  1.4(5) &  5.8 \\
    \hline
  \end{tabular}
\end{table*}

\subsubsection{Fairborn photometry}

The Fairborn light curves show long-term irregular variations, most probably of instrumental origin. These trends were removed by a two-step iterative process. In the first step, we prewhitened each band for the dominant pulsation frequency ($0.652593$\,d$^{-1}$). Next, the residual light curves were fitted with low-order splines season-by-season. Finally, these splines were subtracted from the original light curves, filtering out frequencies below 0.004\,d$^{-1}$, and their aliases from the Fourier spectra.

We analysed the filtered light curves of all four observed Str\"omgren bands ($uvby$) separately, looking for significant frequency components. We accepted only those frequencies that appear in at least three bands with at least 4.0\,$\sigma$ amplitude. Altogether, five frequencies met this criterion. The final frequency fits and S/N calculations were performed using fixed frequency values. These frequencies and their uncertainties were calculated respectively as the average and scatter of the best non-linear frequency fit results for the four passbands. The frequency solution is summarized in Table~\ref{tbl:fairbornfreq}. 

All the frequencies found in the Fairborn data are also detected in the \most\/ light curve, however, the difference between the frequency values of the two data sets usually exceeds their standard errors. On one hand, the frequencies might be Doppler-shifted due to the orbital motion, thus, no exact match is expected for the two data sets. On the other hand, the standard errors of the \most\/ frequencies might underestimate the real uncertainties due to possible unresolved frequency components near the identified ones in the short time-base data.

\subsection{Spectroscopy}

Visual inspection of the mean line profiles already showed that, in accordance with their location in the SPB instability region of the HRD, both components of \oo\ exhibit LPVs.

We looked for significant periodicities in several different data sets derived from the spectroscopic observations. The orbital variations in the relative positions of the lines of the two components (see Sect.~\ref{sect:orbitspect}) force us to analyse the seasons separately. Only the 2008 and 2010 observations are extended enough to permit Fourier analysis based on 193 and 1737 spectra, respectively. We investigated the low-order moments of the cross-correlated line profiles, and also the variations across the whole line profile with the Pixel-by-Pixel (PbP) method, as implemented in the \famias\ software \citep{famias}.

\subsubsection{Moments}

Time series of the 0th--3rd moments ($m_0\,...\,m_3$) and their uncertainties were calculated from the cross-correlated line profiles, using individual S/N values determined empirically by \famias. The continuum was excluded individually from each profile before the moment calculations. The identified frequencies are listed in Table~\ref{tbl:momentfreq}.

The 2008 moment data only allow us to identify the dominant frequency of the star and only in the $m_1$ data. Furthermore, there is no significant variation of the $m_0$ data in any of the two investigated seasons, that is, the EW of the {mean line-profile} is approximately constant over time. The columns of those moments that show no significant variations are omitted from Table~\ref{tbl:momentfreq}.

\begin{table}
  \centering
  \caption{Identified frequencies and their S/N ratios in the 1st--3rd moments ($m_1$, $m_2$, $m_3$) of the cross-correlated line profiles of the spectroscopic observations in the  2008 and 2010 seasons.\label{tbl:momentfreq}}
  \begin{tabular}{cl@{\hskip5mm}c@{\hskip5mm}rrr}
    \hline
                   &             & \multicolumn{3}{c}{S/N in} \\
     ID            & Freq.       &2008& \multicolumn{3}{c}{2010} \\
                   & (d$^{-1}$)  & $m_1$ & $m_1$ & $m_2$ & $m_3$ \\
    \hline
    $f_{\mathrm{mm}\,1}$ & 0.653 & 7.1   & 20.6  & 8.2   & 8.3 \\
    $f_{\mathrm{mm}\,2}$ & 1.676 & --    &  8.6  & --    & 4.7 \\
    $f_{\mathrm{mm}\,3}$ & 1.350 & --    &  8.1  & 4.2   & 9.0 \\
    $f_{\mathrm{mm}\,4}$ & 1.192 & --    &  5.0  & 6.6   & 10.6 \\
    $f_{\mathrm{mm}\,5}$ & 0.020 & --    & --    & 7.0   & 11.9 \\
    $f_{\mathrm{mm}\,6}$ & 0.231 & --    & --    & 3.7   & 5.2 \\
    $f_{\mathrm{mm}\,7}$ & 0.158 & --    & --    & 3.9   & 6.3 \\
    \hline
  \end{tabular}
\end{table}

\begin{figure*}
 \begin{center}
  \includegraphics[width=175mm]{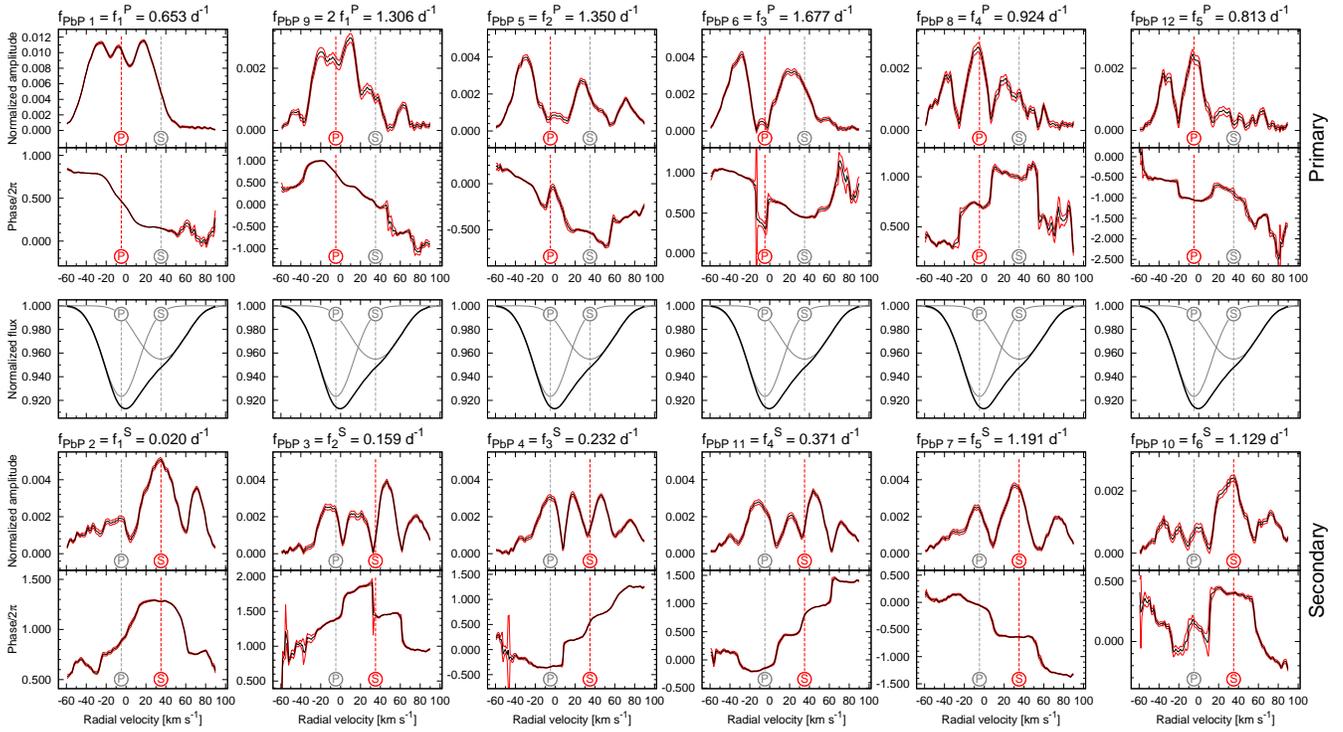}
 \end{center}
 \caption{Amplitude, phase and zero-point profiles of the LPVs with different frequencies for the cross-correlated line profiles of the 2010 observing season. The ZP profiles plotted in the middle row are identical for each frequency. The fitted profiles of the two components are also plotted with thin gray lines. The vertical lines marked with P and S correspond to the location of the centre of the primary and secondary components' lines, respectively. The component to which a given frequency is attributed is highlighted. The thin gray/red lines surrounding the thick (black) middle lines mark the standard error limits of the profiles.\label{fig:pbpprofiles}}
\end{figure*}

\begin{table}
  \centering
  \caption{Frequencies identified by the PbP analysis in the LPV of the mean line profiles of the 2010 spectroscopic observations, and the component to which they are attributed.\label{tbl:pbpfreq}}
  \begin{tabular}{clc}
    \hline
        ID    & Freq. (d$^{-1}$)  & component \\
    \hline
    $f_{\mathrm{PbP}\,1}$ & 0.6528 & Primary \\
    $f_{\mathrm{PbP}\,2}$ & 0.0197 & Secondary \\
    $f_{\mathrm{PbP}\,3}$ & 0.1593 & Secondary \\
    $f_{\mathrm{PbP}\,4}$ & 0.2316 & Secondary \\
    $f_{\mathrm{PbP}\,5}$ & 1.3498 & Primary \\
    $f_{\mathrm{PbP}\,6}$ & 1.6773 & Primary \\
    $f_{\mathrm{PbP}\,7}$ & 1.1906 & Secondary \\
    $f_{\mathrm{PbP}\,8}$ & 0.9246 & Primary \\
    $f_{\mathrm{PbP}\,9}$ & 1.3054 & Primary \\
    $f_{\mathrm{PbP}\,10}$& 1.1291 & Secondary \\
    $f_{\mathrm{PbP}\,11}$& 0.3712 & Secondary \\
    $f_{\mathrm{PbP}\,12}$& 0.8135 & Primary \\
    \hline
  \end{tabular}
\end{table}

\subsubsection{Pixel-by-Pixel Fourier analysis}
\label{sect:pbpanalysis}

We analysed the variations of the line profiles in each wavelength bin using the Pixel-by-Pixel (PbP) method \citep{Schrijvers1997A&AS..121..343S,Telting1997A&A...317..723T,Telting1997A&A...317..742T} as implemented by \famias. Since the amplitude of the LPV can strongly fluctuate across the profile, no straightforward and strict requirements can be set against any periodicity detected by this method to be accepted as significant. Thus, we used Fourier spectra averaged over some sections of the line profile as well as single-pixel spectra to look for strong variations.

The 193 spectra observed in the 2008 season are insufficient to investigate the complex multi-periodic LPVs by PbP analysis. This data set shows only the dominant periodicity with sufficient confidence. Thus, we discuss only our results on the 2010 data in the followings.

We succeeded in identifying 12 variation frequencies in the 2010 season's data. Most of these frequencies are present in other data sets as well, supporting our selection. The two exceptions are $f_{\mathrm{PbP}\,11} = 0.371$\,\cd\ and the second harmonic of the dominant frequency, $f_{\mathrm{PbP}\,9} = 2f_{\mathrm{PbP}\,1} = 1.305$\,\cd. The identified frequencies are listed in Table~\ref{tbl:pbpfreq}.

The results of the PbP analysis, the zero-point, amplitude and phase profiles for each periodicities, are plotted in Fig.~\ref{fig:pbpprofiles}. The zero-point profiles (ZP) are the same for all the frequencies. These are plotted in the middle row of Fig.~\ref{fig:pbpprofiles} multiple times only for easier comparison with the other profiles. We indicated the location of the centre of the primary and secondary components' lines in each panel of Fig.~\ref{fig:pbpprofiles}, marked with P and S. For each profile, one of the two line centres approximate the symmetry axis much better than the other one. Also, the amplitude profiles usually extend towards one side of the blended line profile much more than towards the other side, indicating which component the given periodicity originates from. Based on these morphological features, each frequency can be attributed either to the primary or to the secondary component. We highlighted the corresponding component in each amplitude and phase profile panel in Fig.~\ref{fig:pbpprofiles}. Table~\ref{tbl:pbpfreq} lists the identified frequencies together with the corresponding component. The top and bottom rows of Fig.~\ref{fig:pbpprofiles} show the frequencies belonging to the primary and secondary, respectively. The more-or-less regular shape of the amplitude and phase profiles of these frequencies also support our frequency selection.

In some cases, strong deviation from the symmetry of the profiles is observable. Our tests using the Line Profile Synthesis tool of \famias\ show that the Fourier-parameter profiles of synthetic data might be significantly asymmetric solely due to the time distribution of the observations, even though they are quite numerous, as 1737 observations are available from the 2010 season. The profiles are further distorted by the presence of the companion. As the line profile does not converge to the continuum on the companion's side, significant random amplitudes and phases can be reached in these contaminated regions. These systematic deviations often exceed the standard errors calculated by \famias. This is demonstrated, for example, by the amplitude and phase profiles of the dominant mode above 60\,\kms, where the line of the primary component does not extend (see Fig.~\ref{fig:doubleline}). Here the amplitude deviates from 0 at the 0.001 amplitude level, and the phase shows large fluctuations also. The observed asymmetries might also have real physical origin, for example, can be caused by fast rotation.

Inspecting Fig.~\ref{fig:pbpprofiles}, one has the impression that the locations of the centre of the two components are not quite appropriate. The RVs derived in Sect.~\ref{sect:orbitspect} are apparently offset from the expected axis of symmetry of the amplitude and phase profiles. Both for the primary and for the secondary, a RV shift of about $-2\,...$\,$-3$\,\kms\ seems to be more appropriate. Such a correction would greatly reduce the mass-ratio discrepancy discussed in Sect.~\ref{sect:relphyspar}.

\subsection{Summary and discussion of the frequency analysis}
\label{sect:freqsummary}

The different methods applied to the different data sets to determine periodicities of \oo\ resulted in 11 independent significant frequencies and one harmonic frequency. We also obtained information by the PbP analysis on which frequency originates from which component. Therefore, it is worth to summarize the main results here. A new notation is also introduced taking into account that both components are variable. Thus, we denote frequencies related to the primary and secondary components with $f^{\mathrm{P}}_n$ and $f^{\mathrm{S}}_n$, respectively. The new frequency notation is defined in Table~\ref{tbl:freqsummary}.

\begin{table}
  \centering
  \caption{Summary and new notation of the frequencies found in the different data sets. The last, ``Cross-identification'' column refers to notations used in Tables~\ref{tbl:mostfreq}--\ref{tbl:pbpfreq} ($f_{\mathrm{M}}$ -- \most\ photometry; $f_{\mathrm{Fb}}$ -- Fairborn photometry; $f_{\mathrm{mm}}$ -- line-profile moments; $f_{\mathrm{PbP}}$ -- Pixel-by-Pixel analysis.)}\label{tbl:freqsummary}
  \begin{tabular}{ccll}
    \hline
        ID              & Frequency & \multicolumn{1}{c}{Period} & \multicolumn{1}{c}{Cross-identification} \\
                        & (d$^{-1}$)& \multicolumn{1}{c}{(d)}  \\
    \hline
     $f^{\mathrm{P}}_1$ & 0.653     & \pz1.532  & $f_{\mathrm{M}\,1}$, $f_{\mathrm{Fb}\,1}$, $f_{\mathrm{mm}\,1}$, $f_{\mathrm{PbP}\,1}$ \\
     $2f^{\mathrm{P}}_1$& 1.306     & \pz0.766  & $f_{\mathrm{PbP}\,9}$ \\
     $f^{\mathrm{P}}_2$ & 1.350     & \pz0.741  & $f_{\mathrm{M}\,4}$, $f_{\mathrm{mm}\,3}$, $f_{\mathrm{PbP}\,5}$ \\
     $f^{\mathrm{P}}_3$ & 1.677     & \pz0.596  & $f_{\mathrm{M}\,7}$, $f_{\mathrm{mm}\,2}$, $f_{\mathrm{PbP}\,6}$ \\
     $f^{\mathrm{P}}_4$ & 0.924     & \pz1.082  & $f_{\mathrm{M}\,3}$, $f_{\mathrm{Fb}\,2}$, $f_{\mathrm{PbP}\,8}$ \\
     $f^{\mathrm{P}}_5$ & 0.813     & \pz1.230  & $f_{\mathrm{M}\,5}$, $f_{\mathrm{Fb}\,5}$, $f_{\mathrm{PbP}\,12}$ \\[1.5mm]
     $f^{\mathrm{S}}_1$ & 0.020     & 50.0   & $f_{\mathrm{mm}\,5}$, $f_{\mathrm{PbP}\,2}$ \\
     $f^{\mathrm{S}}_2$ & 0.159     & \pz6.289  & $f_{\mathrm{mm}\,7}$, $f_{\mathrm{PbP}\,3}$ \\
     $f^{\mathrm{S}}_3$ & 0.232     & \pz4.310  & $f_{\mathrm{mm}\,6}$, $f_{\mathrm{PbP}\,4}$ \\
     $f^{\mathrm{S}}_4$ & 0.371     & \pz2.695  & $f_{\mathrm{PbP}\,11}$ \\
     $f^{\mathrm{S}}_5$ & 1.191     & \pz0.840  & $f_{\mathrm{M}\,2}$, $f_{\mathrm{Fb}\,4}$, $f_{\mathrm{mm}\,4}$, $f_{\mathrm{PbP}\,7}$ \\
     $f^{\mathrm{S}}_6$ & 1.129     & \pz0.886  & $f_{\mathrm{M}\,6}$, $f_{\mathrm{Fb}\,3}$, $f_{\mathrm{PbP}\,10}$ \\
    \hline
  \end{tabular}
\end{table}

We did not detect signs of $\beta$ Cep pulsations. There are no significant peaks in the 3--10\,\cd\ frequency range in the Fourier spectra of any of the investigated data sets.

\paragraph*{$\mathbf{f^{\mathrm{P}}_1}$}
-- this is the dominant frequency. It appears in each investigated data set with the exception of the 2008 $m_0$, $m_2$, $m_3$ and 2010 $m_0$ moments. It is quite stable on the time scale of decades, as all the phase deviations shown by the $O-C$ diagram in Fig.~\ref{fig:o-c} can be explained by the orbital light-time variations. Both the $O-C$  and the PbP analysis attribute this frequency to the primary component. Its harmonic, $2f^{\mathrm{P}}_1$, is also detectable in the LPV in the 2010 spectroscopic data.

\begin{figure}
 \begin{center}
  \includegraphics[width=82mm]{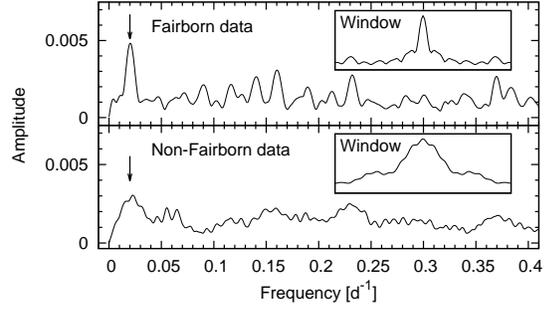}
 \end{center}
 \caption{PbP Fourier analysis of the 20--60\,km\,s$^{-1}$ section of the 2010 Fairborn and non-Fairborn spectroscopic data, after prewhitening for the dominant frequency, $f^{\mathrm{P}}_1$. Inserts show the respective spectral window functions. Both subsets show the periodicity of $f^{\mathrm{S}}_1 = 0.020$\,d$^{-1}$.\label{fig:compare2010subsets}}
\end{figure}

\paragraph*{$\mathbf{f^{\mathrm{S}}_1}$}
-- this is an unusually long periodicity for a SPB star, belonging definitely to the secondary component, according to the PbP analysis. This is the strongest variation of the secondary component, however, only the 2010 spectroscopic data show this frequency. To make sure that this frequency is not an artefact of our data, we analysed separately both the longest homogeneous data set of this season (572 spectra observed in the Fairborn Observatory covering 158 days quite evenly) and the rest of the season's data (1165 spectra covering 218 days). After removing the dominant frequency from the LPVs, the PbP analysis of the 20--60\,km\,s$^{-1}$ section of the profile, where this frequency is quite strong according to Fig.~\ref{fig:pbpprofiles}, shows the peaks of $f^{\mathrm{S}}_1$ for both subsets, as demonstrated in Fig.~\ref{fig:compare2010subsets}. Also the regular, nearly symmetric amplitude and phase profiles obtained by the PbP analysis for this frequency support its intrinsic origin (see Fig.~\ref{fig:pbpprofiles}).

\paragraph*{\bf Low frequencies of the secondary: $\mathbf{f^{\mathrm{S}}_1}$, $\mathbf{f^{\mathrm{S}}_2}$ and $\mathbf{f^{\mathrm{S}}_3}$}
-- these three frequencies of the secondary component are below 0.3\,\cd, thus, they are quite low frequencies for a SPB star. It might be explained with a rotational effect, though. If the secondary is a fast rotator, and these frequencies belong to retrograde modes, then, in the rest frame of the observer, they might be significantly shifted below the usually accepted lower limit of 0.3\,d$^{-1}$ of SPB pulsations. Also, the frequency domain for SPB stars are computed for slow rotation and excitation computation taking rotation into account might explain lower frequencies.

\section{Mode identification}

\subsection{Photometric mode identification}
\label{sect:photmodeid}

We performed photometric mode identification for the five frequencies found in the extended four-colour Str\"omgren photometry obtained in the Fairborn Observatory. The horizontal degrees, $\ell$, of the pulsation modes were identified by matching the observed and theoretically computed amplitude ratios and phase differences in the different passbands. The required non-adiabatic eigenfunctions and eigenfrequencies were computed for modes with $\ell \le 4$ by using the code called {\sc MAD} \citep{Dupret2001A&A...366..166D,Dupret2002A&A...385..563D}. We considered only modes with $\ell \le 4$, since it is quite improbable to detect higher-degree modes in our ground-based data, due to the strong spatial cancellation of these modes.

We selected stellar models in the vicinity of the component values in the parameter space [\teff, \logg, \logL].
Then, we selected theoretical pulsation modes from the stellar models that have frequencies in a 0.2\,\cd\ vicinity of the observed frequency, to allow for frequency shifts introduced by the rotation. Note that the frequency shift of $m\not=0$ modes might be larger than 0.2\,\cd\ even at moderate rotation, thus, we also performed tests with frequency ranges up to 0.6\,\cd. These tests showed no significant differences in the mode-identification results, and the ranking of the modes never changed. The goodness of each individual model is measured by $\chi^2_\mathrm{r}$, characterizing the normalized deviations between the model and the derived physical parameters in the 3-d parameter space and also the deviations between the model and the observations in relative amplitudes and phase differences for three independent Str\"omgren passband pairs ($u$-$v$, $u$-$b$, $u$-$y$). After the set of theoretical modes had been selected for a given observed frequency, we calculated average $\chi^2_\mathrm{r}$ values for each $\ell$ value, and also selected the best-fitting (lowest $\chi^2_\mathrm{r}$) model for each horizontal degree.

\subsubsection{Contamination effect of the companion}

The effect of the companion has to be taken into account when calculating the ratios of the pulsation amplitudes in different passbands. A difference in a particular colour index between the binary components means different contamination in the two passbands. As the observed pulsation amplitude, if expressed in magnitude, is suppressed by the light from the contaminator, the colour index difference distorts the observed amplitude ratios in the investigated passbands. Since we have good estimates of the \teff, \logg\ and \logL\ differences between the two components (see Sect.~\ref{sect:physpar}), we can correct for this effect. We determined the colour-index differences of the two components by linear interpolation in the synthetic Str\"omgren magnitude tables of \cite{Castelli2003IAUS..210P.A20C}, obtaining
\begin{gather*}
(u-v)^\mathrm{P}-(u-v)^\mathrm{S} = \ \,0\fmm000,\\
(u-b)^\mathrm{P}-(u-b)^\mathrm{S} = -0\fmm039,\\
(u-y)^\mathrm{P}-(u-y)^\mathrm{S} = -0\fmm042.
\end{gather*}
\noindent The correction was in most cases less than the 1\,$\sigma$ uncertainty of the amplitude ratio, because of the small temperature and colour difference between the two stars.

The results of our photometric mode identification calculations are summarized in Table~\ref{tbl:photmodeid}.

\begin{table}
  \centering
  \caption{Photometric mode-identification results fitting the amplitude ratios of the multicolour Str\"omgren photometry of \oo.\label{tbl:photmodeid}}
  \begin{tabular}{crrrc}
    \hline
        Freq. (d$^{-1}$)           & $\ell$ & $\langle \chi^2_\mathrm{r} \rangle$ &  $(\chi^2_\mathrm{r})_\mathrm{min}$ & adopted$^{a}$\\
    \hline
        $f^{\mathrm{P}}_1$ = 0.653 &      1 &   1.2  &  0.3 & $\bullet$ \\
                                   &      2 &   5.8  &  1.6 \\
                                   &      3 &  49.0  & 26.3 \\
                                   &      4 &   3.2  &  0.3 \\[1.5mm]
        $f^{\mathrm{P}}_4$ = 0.924 &      1 &   2.2  &  1.7 \\
                                   &      2 &   1.7  &  0.6 & $\circ$ \\
                                   &      3 &   3.9  &  2.2 \\
                                   &      4 &   2.5  &  1.2 \\[1.5mm]
        $f^{\mathrm{P}}_5$ = 0.813 &      1 &   1.0  &  0.5 \\
                                   &      2 &   1.1  &  0.6 \\
                                   &      3 &   1.3  &  0.8 \\
                                   &      4 &   1.1  &  0.6 \\[1.5mm]
        $f^{\mathrm{S}}_5$ = 1.191 &      1 &   1.2  &  0.8 & $\circ$ \\
                                   &      2 &   1.8  &  0.8 & $\circ$ \\
                                   &      3 &   7.3  &  1.4 \\
                                   &      4 &   2.5  &  0.9 \\[1.5mm]
        $f^{\mathrm{S}}_6$ = 1.129 &      1 &   1.3  &  0.9 & $\circ$ \\
                                   &      2 &   5.4  &  1.3 \\
                                   &      3 &  10.7  &  3.1 \\
                                   &      4 &   5.1  &  1.1 \\
    \hline
    \multicolumn{5}{l}{$^a$: $\bullet$ -- certain identification,}\\
    \multicolumn{5}{l}{\ \ \ \,$\circ$ -- ambiguous identification.}\\
  \end{tabular}
\end{table}

\subsubsection{Discussion of the photometric mode-identification results}

The photometric mode identification of the dominant frequency as an $\ell=1$ mode is quite certain, in accordance with the previous result of \cite{decat07}, which was based on different data sets and different models. Although, there is an $\ell=4$ solution with $\chi^2_\mathrm{r}=0.3$ goodness, the $\ell=4$ horizontal degree is rejected in this case, since it is quite improbable that the by far strongest brightness variations are caused by such a high-degree mode.
 
It is also interesting to note that the $\ell=3$ solutions appear to be the least probable for each frequency.

There are only marginal differences between the goodness of the different-degree fits in the case of the weakest signal, $f^{\mathrm{P}}_5$, because the error ranges of the relative amplitudes and phase differences are quite large in this case.

\subsection{Spectroscopic mode identification}

The number of available spectra and the partial separation of the line profiles of the two components of \oo\ allow spectroscopic mode identification for the 2010 data only. We fitted the amplitude and phase profiles shown in Fig.~\ref{fig:pbpprofiles} with theoretical profiles utilizing the Fourier Parameter Fit (FPF) method, as implemented in \famias\ \citep{famias}.

The blending of the line profiles of the two components and the discrepancy in their RVs, as mentioned in Sect.~\ref{sect:pbpanalysis} and demonstrated in Fig.~\ref{fig:pbpprofiles}, makes the use of the ZP profile difficult and ambiguous in the mode-identification fitting process. Since the ZP profile of the binary is the superposition of two ZP profiles of the two components, to fit the ZP profile of the investigated component with the FPF method, the ZP profile of the companion has to be removed in advance. We accomplished this by subtracting one of the profiles fitted to the time-averaged cross-correlated line profiles in Sect.~\ref{sect:orbitspect} (shown in Fig.~\ref{fig:doubleline}) from all the cross-correlated profiles. 

The distortion of the amplitude and phase profiles of the different frequencies, caused by the companion (see Sect.~\ref{sect:pbpanalysis} for discussion) introduces further uncertainty in the mode-identification process. To investigate the ambiguity caused by the different uncertainties, we conducted the mode identification of each frequency by fitting different parts of the profiles and either fitting or disregarding the ZP profiles. The fitting process was applied in the following four different ways:

\begin{itemize}
 \item[$\bullet$] {\bf APf}: The amplitude and phase profiles were fitted (AP fit) to the full line profile: in the $\{-60\,...\,50\,\mathrm{km\,s^{-1}}\}$ and in the $\{-20\,...\,90\,\mathrm{km\,s^{-1}}\}$ range for the primary and secondary, respectively.
 \item[$\bullet$] {\bf ZAPf}: Similar to the APf, but the zero-point profile was also fitted (ZAP fit) in the whole profile range.
 \item[$\bullet$] {\bf APh}: AP fit to that half of the line profile that is least affected by the companion: in the $\{-60\,...\,-5\,\mathrm{km\,s^{-1}}\}$ and in the $\{35\,...\,90\,\mathrm{km\,s^{-1}}\}$ range for the primary and secondary, respectively.
 \item[$\bullet$] {\bf ZAPh}: ZAP fit to the same half of the line profile.
\end{itemize}

We used the fixed values of $R^\mathrm{P} = 2.9\pm1.0\,R_{\odot}$ and $R^\mathrm{S} = 2.45\pm1.1\,R_{\odot}$ radii (calculated from \teff\ and $L$ using the Stefan--Boltzmann equation, as expressed in, for example, \citealt[][eq. 3]{ipm}), $M^\mathrm{P} = 4.6\,M_{\odot}$ and $M^\mathrm{S} = 4.2\,M_{\odot}$ masses, [Fe/H]\,$= -0.3$\,dex metallicity and \teff\ and \logg\ as given in Sect.~\ref{sect:physpar} for modelling the LPV. Our tests show that a difference of $0.1$\,$M_{\odot}$ introduces only negligible changes in the best fitting stellar and pulsational parameters during the mode identification.

\subsubsection{Primary -- identification of the dominant mode, $f^{\mathrm{P}}_1$,  inclination and rotation}
\label{sect:spmodeiddominantmode}

\begin{figure}
 \begin{center}
  \includegraphics[width=88mm]{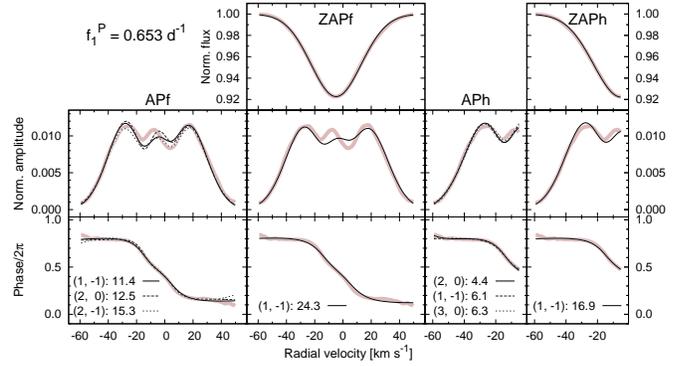}
 \end{center}
 \caption{Spectroscopic mode-identification results for fitting the Fourier-parameter profiles of the dominant frequency, $f^{\mathrm{P}}_1$. The observations are plotted with thick gray/light brown lines. The standard errors of the observations are not indicated, because their values are lower than the thickness of these lines. The fitted models are plotted with different-style thin lines. The $(\ell, m)$ and the corresponding chi-square values are given for each model. Note that the slight asymmetry in the fitted models is due to the low number of phase points (10, non-adjustable) used by \famias\ to model the LPV.\label{fig:spmodeidfP1}}
\end{figure}

We used the mode identification of the dominant frequency, originating from the primary, to derive the inclination and rotation of this component. The results of fitting the dominant mode are summarized in the first section of Table~\ref{tbl:spmodeidfullprim} and in Fig.~\ref{fig:spmodeidfP1}. Here we show the best fitting modes and every other modes with a goodness of fit within 150 per-cent of the best one for each of the four fitting methods.

Our results show almost univocally that the dominant mode is $(\ell, m) = (1, -1)$. Note that we use the same convention as \famias\ for the sign of the azimuthal order, $m$, that is, a negative value denotes a retrograde mode. Only the APh fit resulted in a better goodness for the (2, 0) mode, while the $(1, -1)$ mode gives the best fit with all the other methods. The $\ell=2$ and $\ell=3$ modes can also be rejected on the basis of the photometric mode identification of the dominant frequency (see Table~\ref{tbl:photmodeid}).

The $(1, -1)$ solutions of all the four methods are also consistent in terms of stellar parameters, as shown in Table~\ref{tbl:spmodeidfullprim}. We accept the results for the ZAPh fitting method, because this method uses all three profiles, but takes into account only their left halves, which are almost unaffected by the presence and variations of the companion and by the shape of the fitted and subtracted secondary ZP profile.

\begin{figure}
 \begin{center}
  \includegraphics[width=82mm]{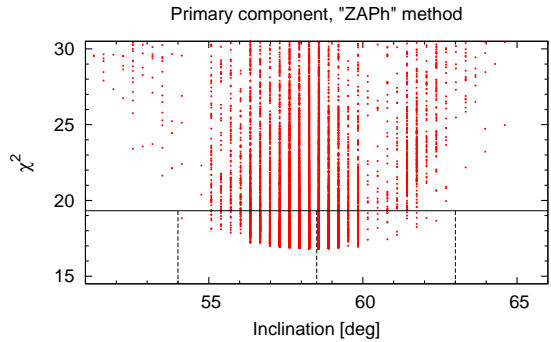}
 \end{center}
 \caption{Goodness of fit vs. inclination for different combination of trial parameters according to the ZAPh fitting method of the dominant frequency with $(\ell, m) = (1, -1)$ mode. The 95 per-cent confidence interval of the inclination of the primary component is marked (where $\chi^2_\mathrm{r} < 1.15 (\chi^2_\mathrm{r})_\mathrm{min}$).\label{fig:inclprimconf}}
\end{figure}

The 95 per-cent confidence intervals for the stellar parameters were determined by inspecting the variation of the minimum of $\chi^2_\mathrm{r}$ with the given parameter. This procedure is demonstrated in Fig.~\ref{fig:inclprimconf} for the inclination. The obtained parameters are: $i^\mathrm{P} = 59\degr\pm 5\degr$, $(v\sin i)^\mathrm{P} = 21.5 \pm 1.5$\,\kms, fitted mean radial velocity: $\Delta Z = -5\pm1$\,\kms. 

Now that the inclination, the projected rotational velocity and the radius of the primary are determined, we can estimate the equatorial rotational velocity: $v^\mathrm{P}_\mathrm{eq} = 25 \pm 2$\,\kms, the rotation period and frequency of this component: $P^\mathrm{P}_\mathrm{rot} = 5.9 \pm 2$\,d, $f^\mathrm{P}_\mathrm{rot} = 0.17 \pm 0.06$\,\cd. This confirms that the primary is a relatively slow rotator, as earlier investigations already suggested. The equatorial rotational velocity is about 4 per-cent of the critical break-up velocity (550 \kms) of the primary component. The lowest pulsation frequency of the primary component is the dominant one. Its value in the co-rotating frame is $f^\mathrm{P}_\mathrm{1\,corot} = f^\mathrm{P}_1 - m f^\mathrm{P}_\mathrm{rot} \approx 0.823$\,\cd, which is significantly larger than the rotational frequency. The spin parameter ($\eta = 2 f_\mathrm{rot}/f_\mathrm{puls\,corot}$) of this mode is $0.41\pm0.15$. Consequently, the first-order approximation of the Coriolis force used by \famias\ for LPV modelling is a-posteriori justified for the primary component.

The inclination angle of complete cancellation (IACC) and the inclination angle of least cancellation (IALC) for an $(\ell, m)= (1, -1)$ mode is $0\degr$ and $90\degr$, respectively \citep[][table~3.1]{DeRidder_PhD}. Therefore, the high inclination yielded by the mode identification is consistent with this frequency being the dominant one in each photometric data set.

\subsubsection{Secondary -- identification of $f^{\mathrm{S}}_2, f^{\mathrm{S}}_3$ and $f^{\mathrm{S}}_4$, inclination and rotation}
\label{sect:modeidsecondary}

Referring to Fig.~\ref{fig:pbpprofiles}, we can see that three frequencies of the secondary, $f^{\mathrm{S}}_2, f^{\mathrm{S}}_3$ and $f^{\mathrm{S}}_4$, show quite similar amplitude and phase variations across the line profile: They all have four amplitude bumps across the line profile, located symmetrically around the central axis of the secondary component, and there is about 1/2 phase shift between the adjacent bumps, while the phase within the bumps is approximately constant. Note that the phase is normalized to 1, thus, phase shifts of any integer numbers are equivalent with each other. We used these three frequencies together to derive the inclination and rotation parameters of the secondary component similarly to what we did for the primary by fitting the profiles of the dominant frequency. Model fits to the individual frequency's profiles show that all three are most probably $(\ell, m) = (2, -2)$ modes (see Sect.~\ref{sect:spmodeidsum} and Table~\ref{tbl:spmodeidfullsec}). We also performed simultaneous fits of the stellar and pulsation parameters with \famias, assuming that all of these three modes have the same $(\ell, m)$ values, with all the four fitting methods. The results of these fits are summarized in Table~\ref{tbl:spmodeidsecondary}. In this table, we list the best-fitting $(\ell, m)$ modes and any other fit results within 150 per-cent goodness of the best ones with all four fitting methods. Table~\ref{tbl:spmodeidsecondary} shows that, assuming the same $(\ell, m)$ values for the three investigated modes, they are $(\ell, m) = (2, -2)$ ones with the highest probability.

We again accept the stellar parameter results obtained with the ZAPh method of this simultaneous fit, for the same reason as for the primary: this method uses information of all three profiles, but now only from the right-hand half, which is less affected by the companion, and by the actual ZP profile shape fitted for the primary and subtracted from the line profiles. The accepted stellar parameters and their 95 per-cent confidence intervals are: $i^\mathrm{S} = 20\degr^{+7}_{-5}$, $(v\sin i)^\mathrm{S} = 35 \pm 4$\,\kms, $\Delta Z = 35\pm2$\,\kms.

The rotation parameters of the secondary, based on these results, are: $v^\mathrm{S}_\mathrm{eq} = 100 \pm 30$\,\kms, $P^\mathrm{S}_\mathrm{rot} = 1.2 \pm 0.6$\,d, $f^\mathrm{S}_\mathrm{rot} = 0.8 \pm 0.4$\,\cd. That is, according to the mode-identification results of these three frequencies, the secondary is a fast rotator. The equatorial rotational velocity is about 18 per-cent of the critical break-up velocity (570 \kms) of the secondary component.

The calculation of the spin parameter of a retrograde mode at fast rotation is ambiguous, because the pulsation frequency in the corotating frame itself is ambiguous. We cannot be sure whether what we see is really a retrograde propagating pattern on the stellar surface ($f_\mathrm{rot} < f_\mathrm{puls\,corot}$), or the rotation is so fast that the speed of rotation exceeds the speed of propagation of the mode on the stellar surface. In the latter case, we actually observe a prograde moving pattern even though we are dealing with a retrograde mode in the co-rotating frame ($f_\mathrm{rot} > f_\mathrm{puls\,corot}$). For $m=-2$ modes, the rotation limit between the two cases is just at $\eta = 1$ (because $f_\mathrm{rot} = f_\mathrm{puls\,corot}$). Thus, we calculated two possible values of the spin parameter for each of these three frequencies, using two possible pulsation frequencies in the co-rotating frame: $f_\mathrm{puls\,corot} = | f_\mathrm{puls\,obs} \pm 2 f_\mathrm{rot}|$. The difference between the two cases is the largest for the $(2, -2)$ mode with the highest frequency, $f^{\mathrm{S}}_4$. For this mode, the spin parameter is either 0.8 or 1.3, that is, we are either just within or somewhat outside the claimed validity range of the first-order approximation of the Coriolis force used in \famias\ \citep{famias}.

In this case, the FPF mode-identification results obtained by \famias\ are somewhat questionable. However, there are two arguments that support the validity of our mode-identification results of these three frequencies:
\begin{itemize}
\item[$\bullet$] The three frequencies investigated in this section are not detected in the photometric light curves. This is consistent with their spectroscopic mode-identification, since the IACC and IALC for an $(\ell, m) = (2, -2)$ mode is $0\degr$ and $90\degr$, respectively. The derived low inclination angle would cause almost complete photometric cancellation.
\item[$\bullet$] The frequencies of these modes, and especially the frequency of $f^{\mathrm{S}}_1 = 0.02$\,\cd, are rather low for a SPB variable. Negative $m$ values of these modes together with fast rotation would explain the significant shift towards the low frequency domain of these modes. The derived $m = -2$ azimuthal order of $f^{\mathrm{S}}_2, f^{\mathrm{S}}_3$ and $f^{\mathrm{S}}_4$ are consistent with such an explanation. No other $m < 0$ solution results in similarly good fit for the PbP profiles of these three modes.
\end{itemize}

As \cite{Townsend2003MNRAS.343..125T} pointed out, the retrograde modes are more affected by the fast rotation, and these modes are confined to a waveguide near the equator. Such a scenario would mean worse spectroscopic visibility at low inclination angles. Thus, we speculate that the inclination angle we derived should most probably be increased somewhat. However, if the inclination is increased, then the equatorial rotation becomes slower, while the waveguide effect weakens. Therefore, the possible correction in the inclination, if the fast rotation were taken into account, would be small, probably less than $10\degr$. The equatorial rotational velocity for $i=30\degr$ would be 70\,\kms, and the rotation frequency would be 0.55\,\cd\ in this case.

\subsubsection{Mode identification of the other frequencies}
\label{sect:modeidrest}

\begin{figure*}
 \begin{center}
  \includegraphics[width=85mm]{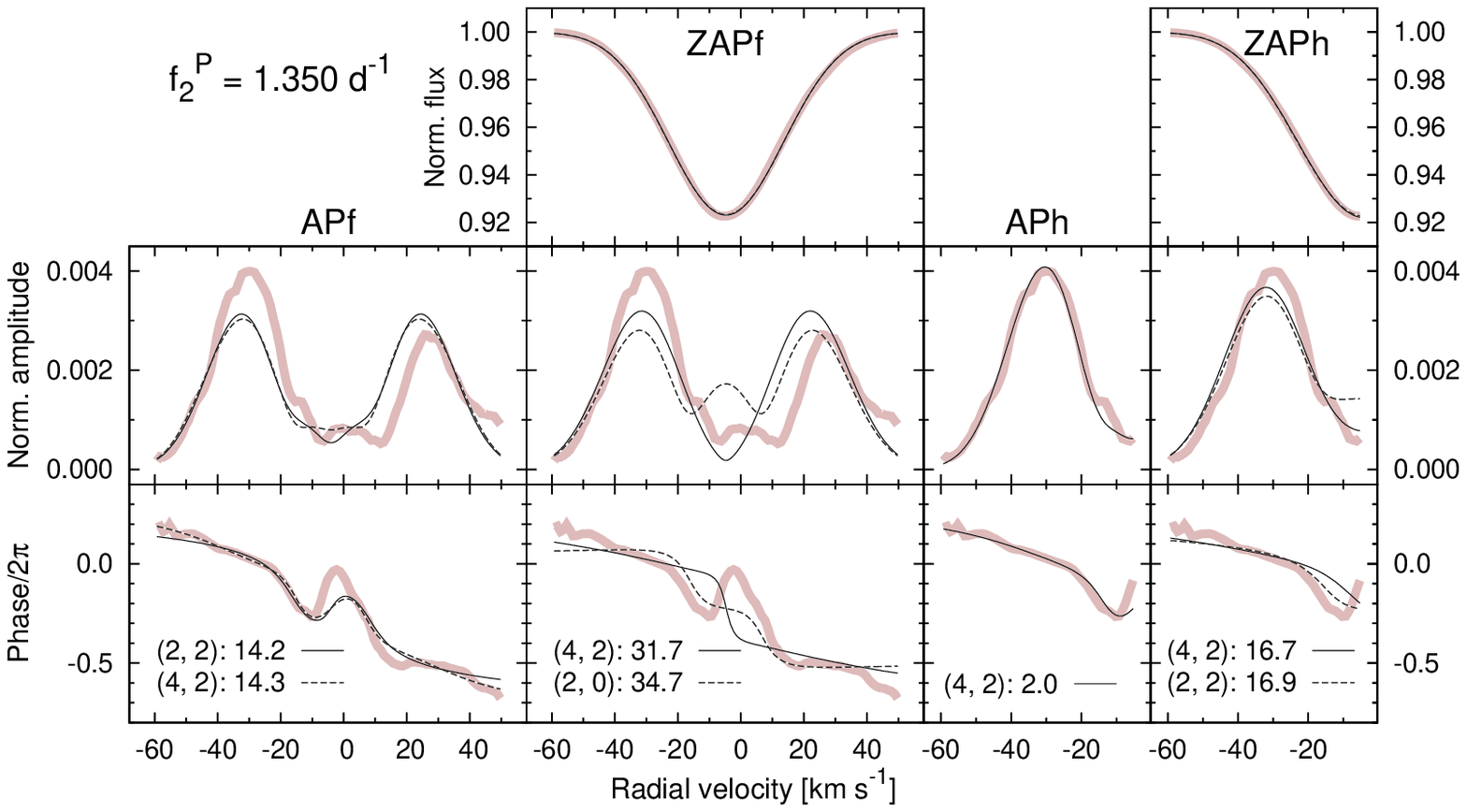}
  \includegraphics[width=85mm]{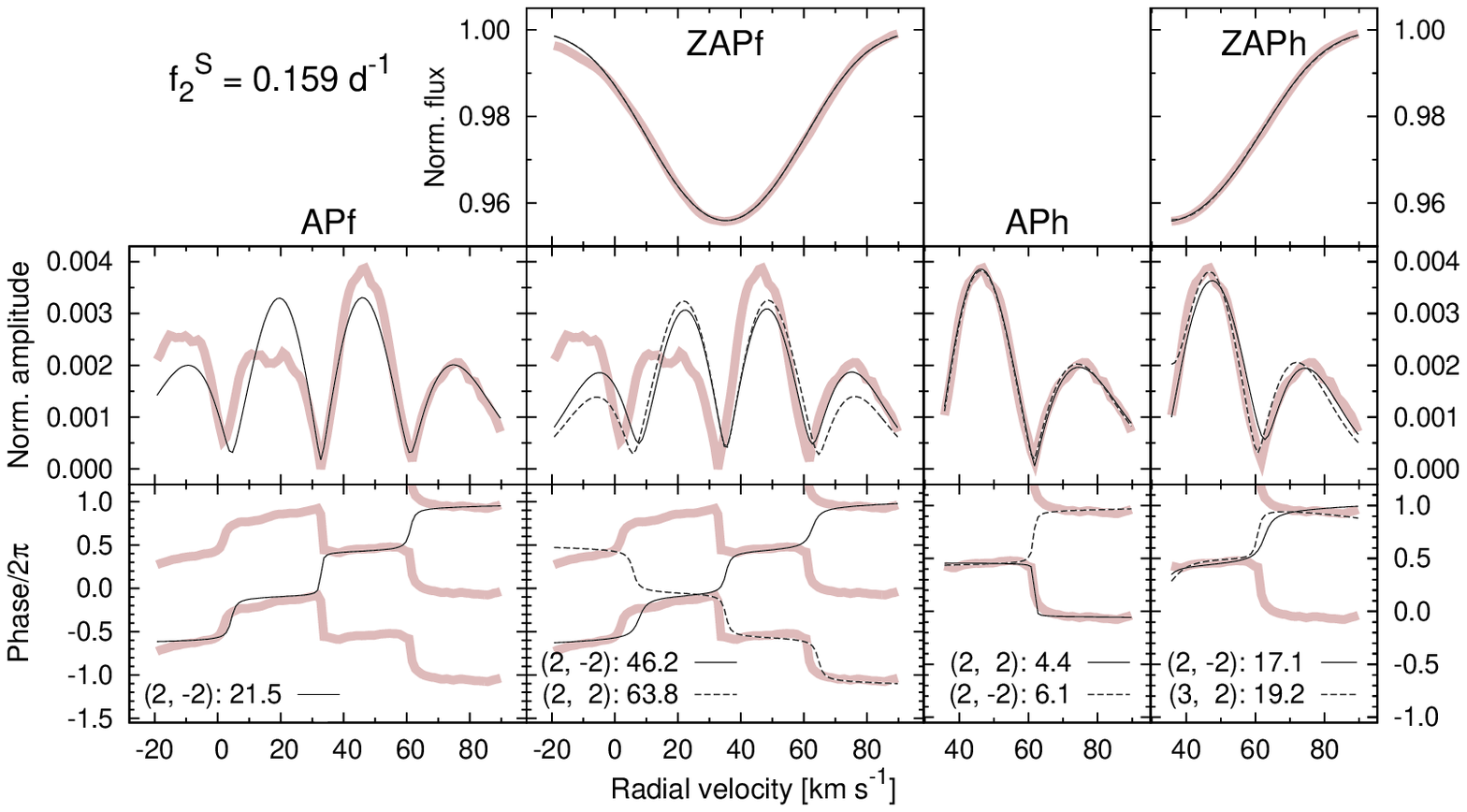}

  \includegraphics[width=85mm]{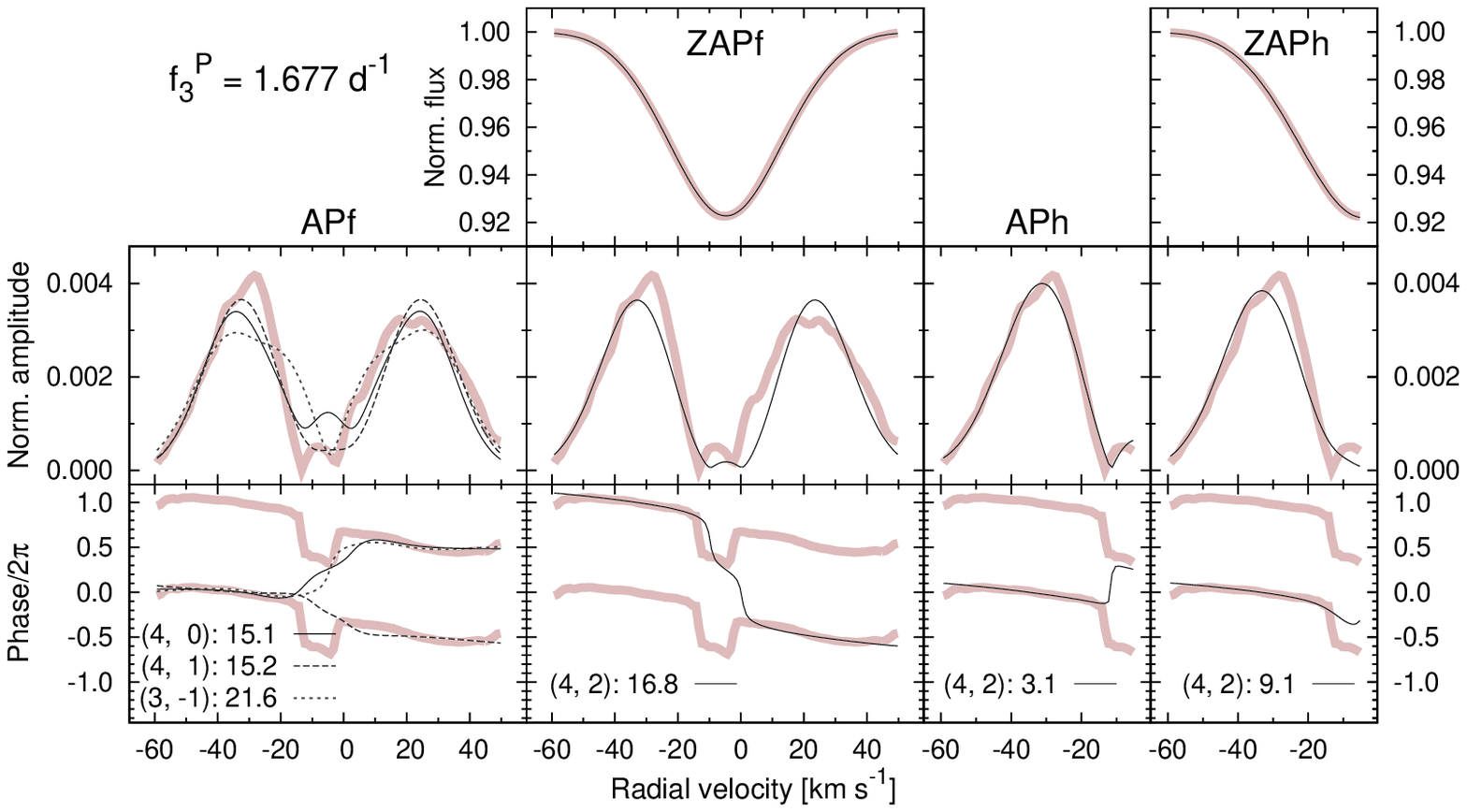}
  \includegraphics[width=85mm]{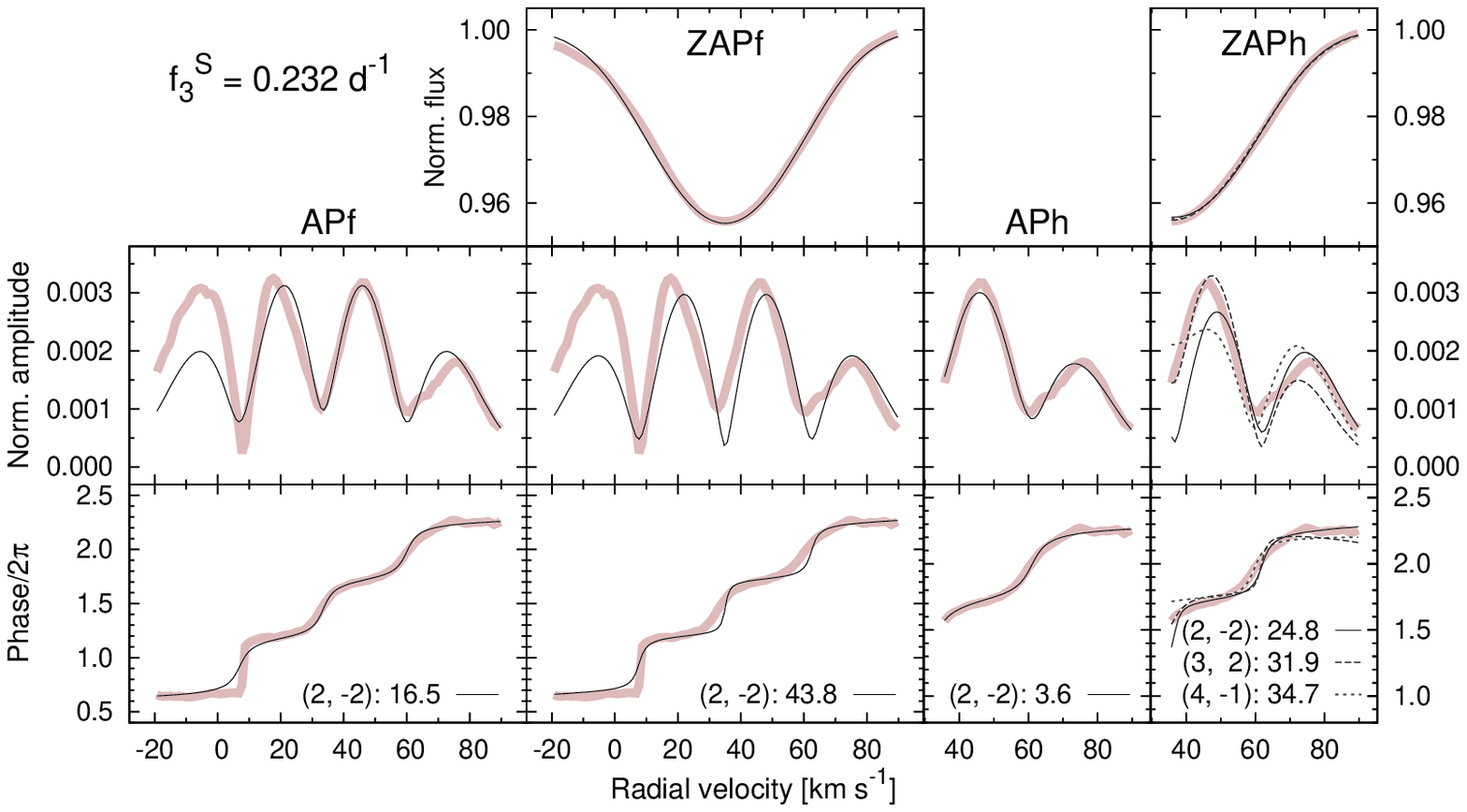}

  \includegraphics[width=85mm]{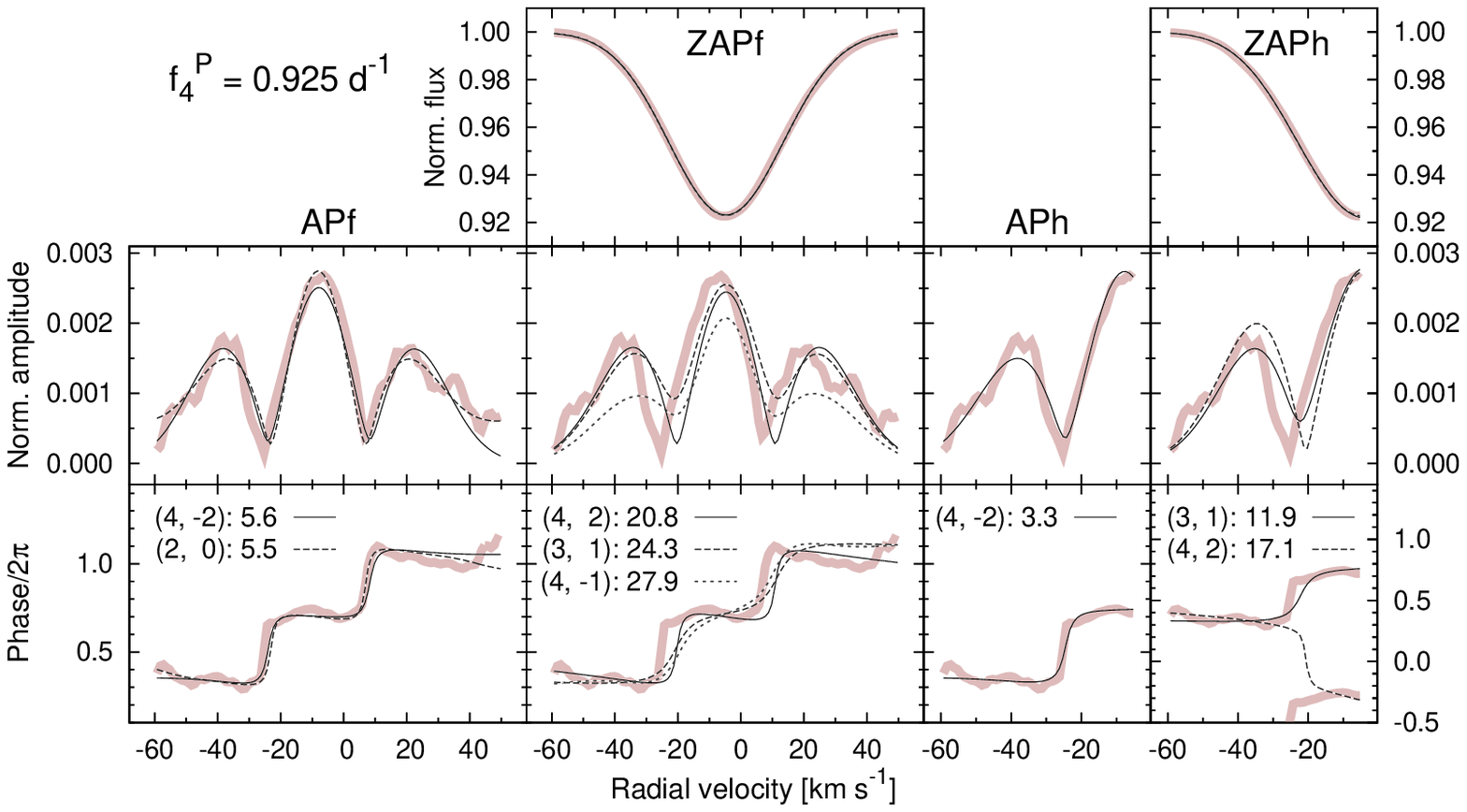}
  \includegraphics[width=85mm]{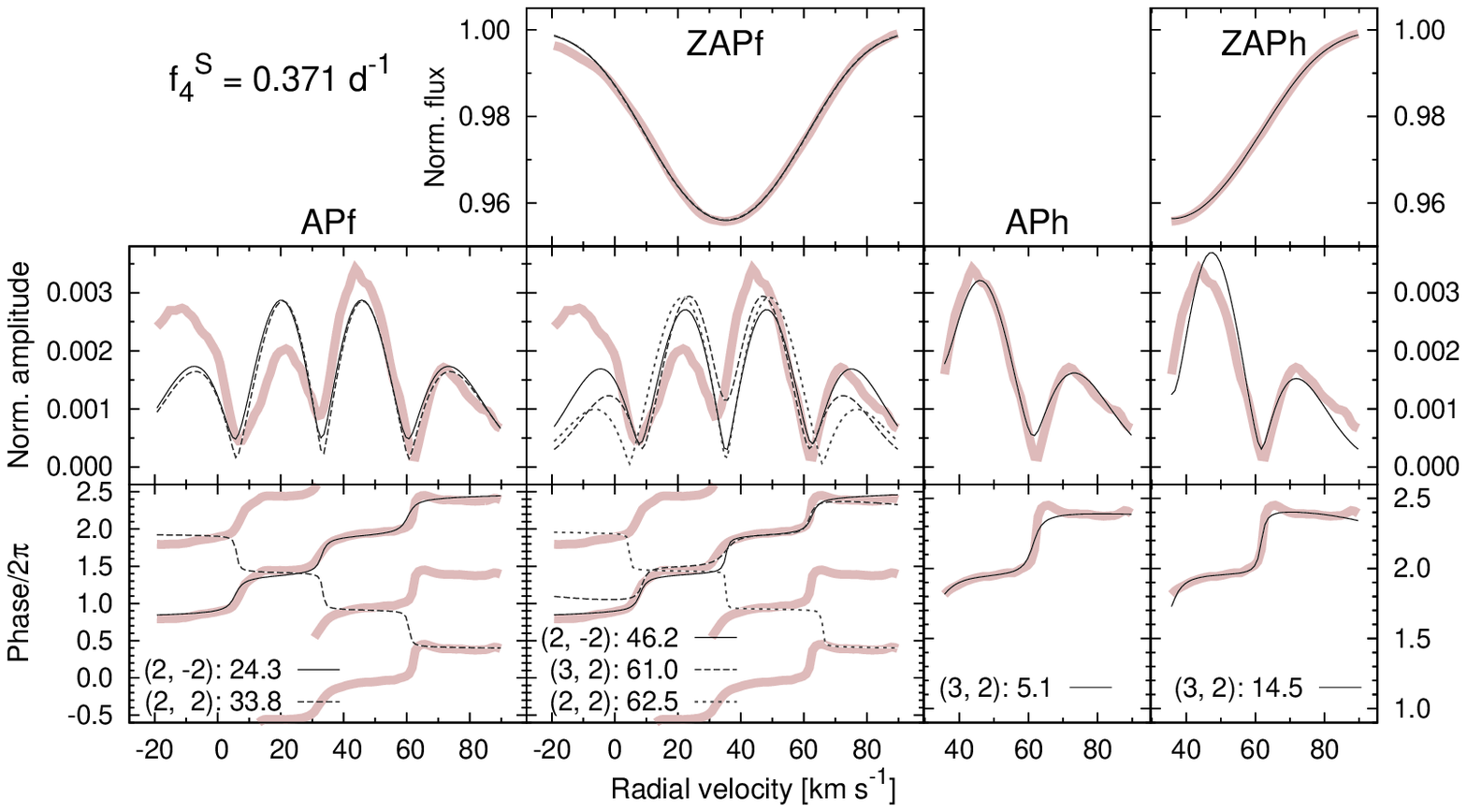}

  \includegraphics[width=85mm]{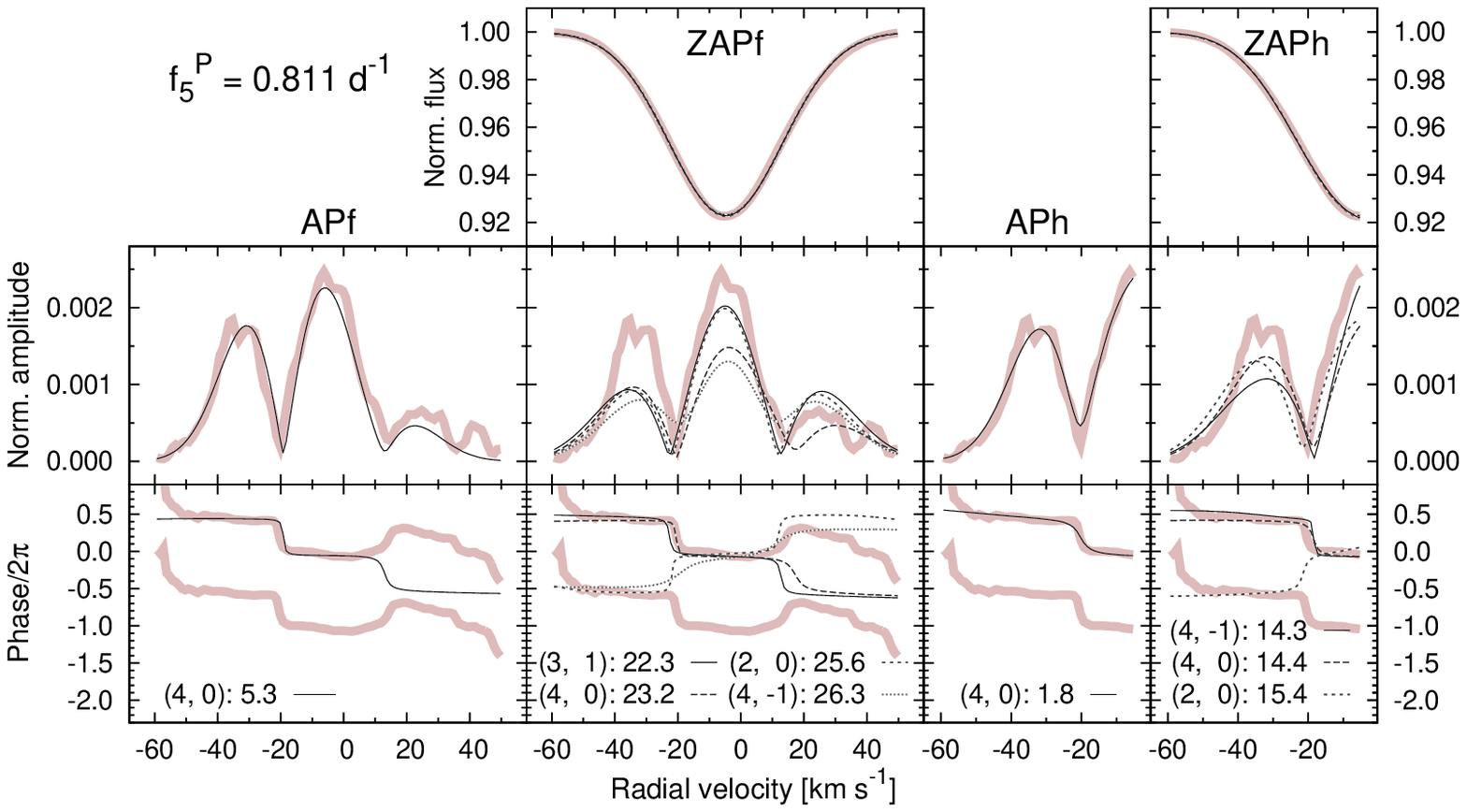}
  \includegraphics[width=85mm]{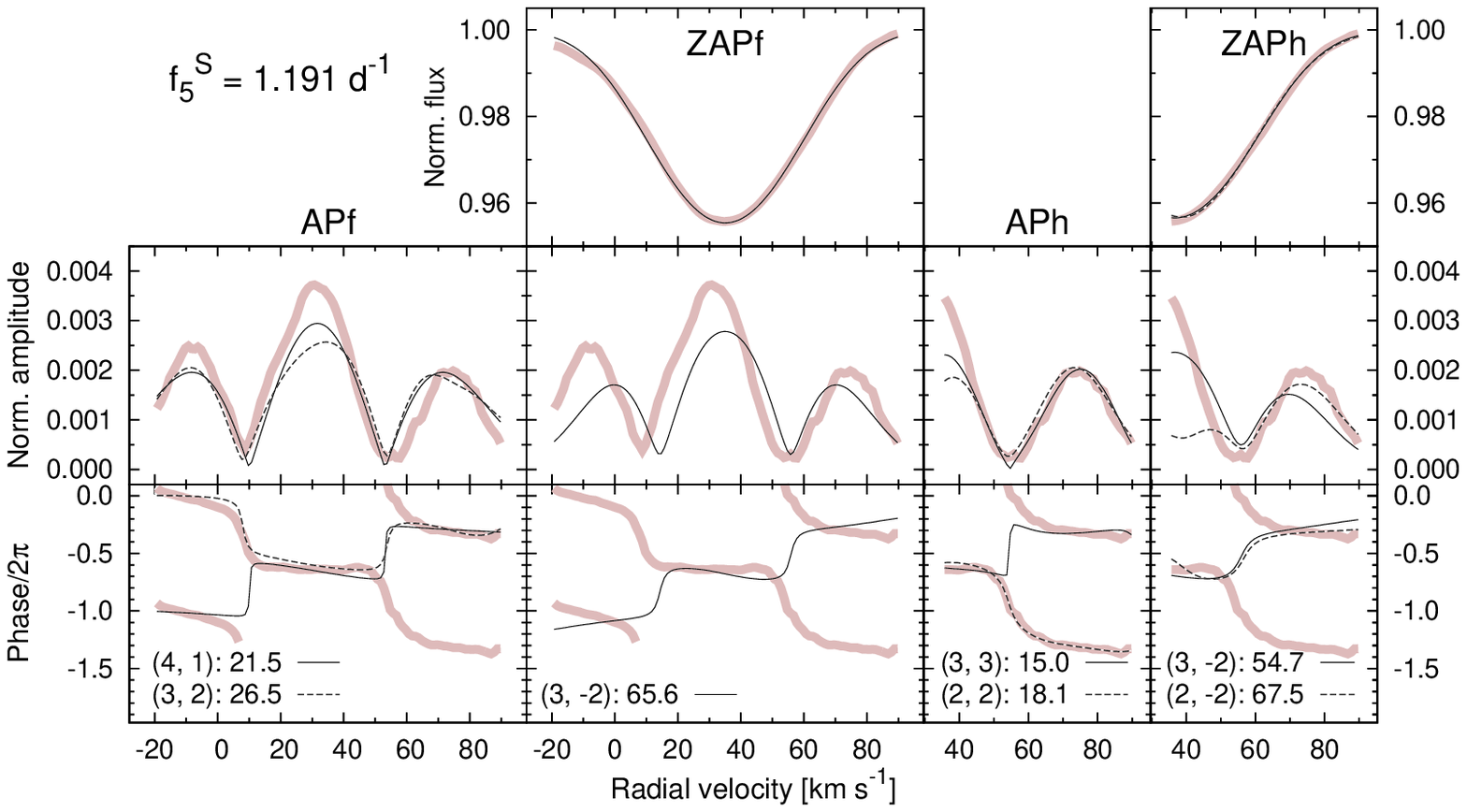}

  \includegraphics[width=85mm]{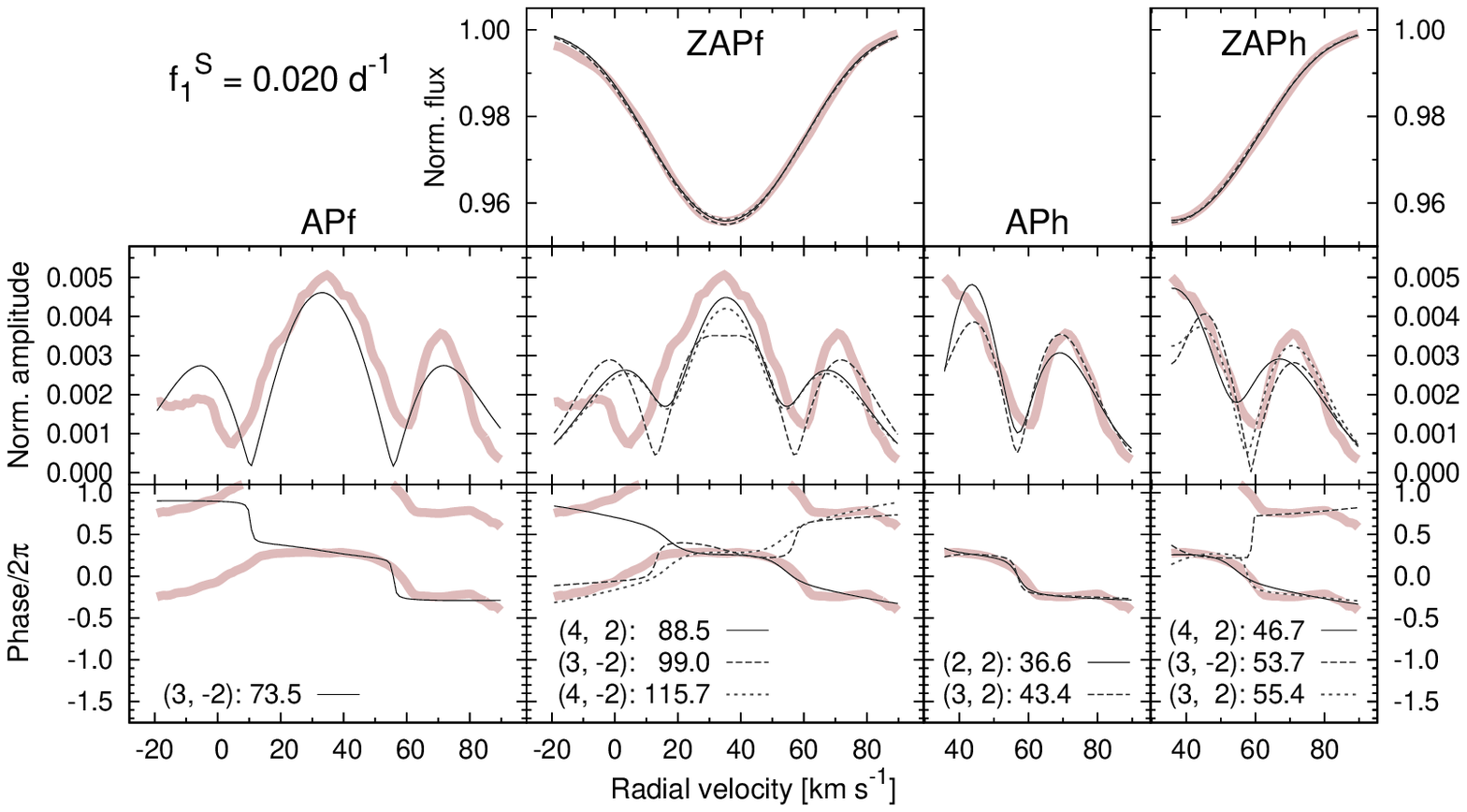}
  \includegraphics[width=85mm]{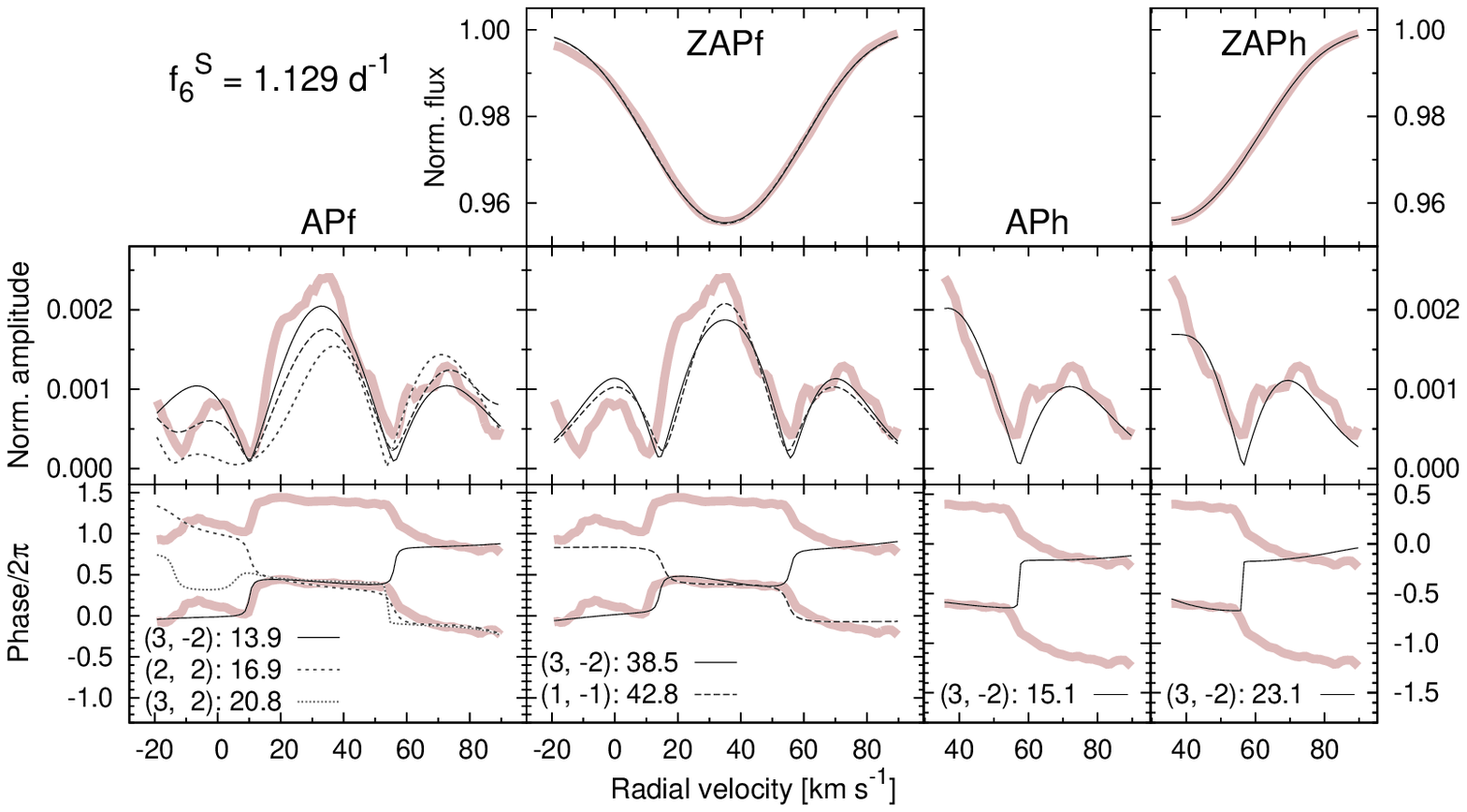}

  \end{center}
 \caption{The same as Fig.~\ref{fig:spmodeidfP1}, but for $f^{\mathrm{P}}_2\,...\,f^{\mathrm{P}}_5$ and $f^{\mathrm{S}}_1\,...\,f^{\mathrm{S}}_6$.\label{fig:spm}}
\end{figure*}

We performed the mode identification of the rest of the independent frequencies listed in Table~\ref{tbl:freqsummary} with the FPF method using all four fitting methods. The results of these fits are given in Tables~\ref{tbl:spmodeidfullprim}~and~\ref{tbl:spmodeidfullsec} for the frequencies of the primary and the secondary, respectively, and are plotted in Fig.~\ref{fig:spm}. We accepted only those solutions that have inclinations within 2 times the 95 per-cent confidence interval of the derived inclination of the corresponding component (49$\degr$--69$\degr$ for the primary and 10$\degr$--35$\degr$ for the secondary), that is, we restricted the searching interval of the inclination within \famias\ to these ranges when running the optimisation. For each frequency, we list the modes that best fit the Fourier-parameter profiles and also those that have a goodness of fit within 150 per-cent of the best fitting mode.

\begin{table*}
  \centering
  \caption{Spectroscopic mode-identification results for fitting the Fourier-parameter profiles of the independent frequencies of the primary component of \oo.\label{tbl:spmodeidfullprim}}
  \begin{tabular}{lcrrrrrrc}
    \hline
    Method      & $(\ell, m)$ & $\chi^2_\mathrm{r}$ & $i$ & \vsini& $\sigma^{a}$ & $\Delta Z^{b}$ & $A(v)^{c}$ & adopted$^{d}$\\
                &             &                     &(deg)& \kms  & \kms         &    \kms        & \kms       \\
    \hline
    \multicolumn{9}{c}{$f^{\mathrm{P}}_1 = 0.653$\,\cd}\\[1.5mm]
    APf         &  (1, -1) &                11.4 &  67 &     22.2 &        12.4  &           -5.5 &       1.8  & $\bullet$\\
                &  (2,  0) &                12.5 &  47 &     18.3 &        11.2  &           -5.8 &       1.3 \\
                &  (2, -1) &                15.3 &   8 &     17.6 &        14.9  &           -5.3 &       4.0 \\[1.5mm]
       
    ZAPf        &  (1, -1) &                24.3 &  55 &     19.9 &        14.2  &           -4.7 &       1.5  & $\bullet$\\[1.5mm]

    APh         &  (2,  0) &                 4.4 &  51 &     16.4 &        12.2  &           -5.0 &       1.8 \\
                &  (1, -1) &                 6.1 &  52 &     21.2 &        14.3  &           -3.1 &       1.0  & $\bullet$\\
                &  (3,  0) &                 6.3 &  35 &     15.8 &        13.8  &           -4.4 &       0.6 \\[1.5mm]

    ZAPh        &  (1, -1) &                16.9 &  59 &     21.4 &        13.3  &           -4.9 &       1.5  & $\bullet$\\
    \hline
    \multicolumn{9}{c}{$f^{\mathrm{P}}_2 = 1.350$\,\cd}\\[1.5mm]
    APf         &  (2,  2) &                14.2 &  61 &     27.4 &        11.3  &           -4.0 &       1.0  & $\circ$\\
                &  (4,  2) &                14.3 &  69 &     18.3 &         9.5  &           -4.0 &       1.9  & $\circ$\\[1.5mm]

    ZAPf        &  (4,  2) &                31.7 &  65 &     22.0 &        12.8  &           -4.7 &       1.1  & $\circ$\\
                &  (2,  0) &                34.7 &  48 &     20.8 &        14.2  &           -4.7 &       1.8 \\[1.5mm]

    APh         &  (4,  2) &                 2.0 &  68 &     20.3 &         9.8  &           -5.4 &       1.6  & $\circ$\\[1.5mm]

    ZAPh        &  (4,  2) &                16.7 &  66 &     24.2 &        11.8  &           -4.2 &       1.0  & $\circ$\\
                &  (2,  2) &                16.9 &  49 &     23.9 &        13.2  &           -4.2 &       0.6  & $\circ$\\
    \hline
    \multicolumn{9}{c}{$f^{\mathrm{P}}_3 = 1.677$\,\cd}\\[1.5mm]
    APf         &  (4,  0) &                15.1 &  49 &     19.4 &        12.0  &           -5.0 &       1.8 \\
                &  (4,  1) &                15.2 &  56 &     27.6 &        12.0  &           -4.0 &       0.8 \\
                &  (3, -1) &                21.6 &  63 &     20.3 &        13.1  &           -4.5 &       4.9 \\[1.5mm]

    ZAPf        &  (4,  2) &                16.8 &  66 &     22.7 &        12.0  &           -4.7 &       2.2  & $\bullet$\\[1.5mm]

    APh         &  (4,  2) &                 3.1 &  66 &     20.4 &        12.0  &           -4.3 &       2.4  & $\bullet$\\[1.5mm]

    ZAPh        &  (4,  2) &                 9.1 &  66 &     24.1 &        11.4  &           -4.4 &       2.0  & $\bullet$\\
    \hline
    \multicolumn{9}{c}{$f^{\mathrm{P}}_4 = 0.925$\,\cd}\\[1.5mm]
    APf         &  (4, -2) &                 5.5 &  69 &     19.0 &        13.5  &           -8.0 &       0.8 \\
                &  (2,  0) &                 5.6 &  59 &     16.9 &        10.4  &           -8.0 &       5.7 \\[1.5mm]

    ZAPf        &  (4,  2) &                20.8 &  51 &     18.9 &        14.9  &           -4.7 &       0.2 \\
                &  (3,  1) &                24.3 &  58 &     17.0 &        15.2  &           -4.7 &       0.6 \\
                &  (4, -1) &                27.9 &  54 &     21.3 &        12.6  &           -4.7 &       1.4 \\[1.5mm]

    APh         &  (4, -2) &                 3.3 &  69 &     20.6 &        13.5  &           -7.8 &       1.2 \\[1.5mm]

    ZAPh        &  (3,  1) &                11.9 &  55 &     23.3 &        13.6  &           -3.7 &       0.3 \\
                &  (4,  2) &                17.1 &  51 &     22.8 &        13.4  &           -4.4 &       0.1 \\
    \hline
    \multicolumn{9}{c}{$f^{\mathrm{P}}_5 = 0.813$\,\cd}\\[1.5mm]
    APf         &  (4,  0) &                 5.3 &  67 &     13.6 &        12.2  &           -7.8 &       0.6  & $\circ$\\[1.5mm]

    ZAPf        &  (3,  1) &                22.3 &  55 &     13.5 &        16.2  &           -4.7 &       0.4 \\
                &  (4,  0) &                23.2 &  67 &     17.0 &        14.3  &           -4.7 &       0.8  & $\circ$\\
                &  (2,  0) &                25.6 &  55 &     12.1 &        15.7  &           -4.7 &       1.4 \\
                &  (4, -1) &                26.3 &  64 &     19.3 &        14.7  &           -4.7 &       0.5 \\[1.5mm]

    APh         &  (4,  0) &                 1.8 &  69 &     23.7 &        10.8  &           -2.4 &       0.7  & $\circ$\\[1.5mm]

    ZAPh        &  (4, -1) &                14.3 &  62 &     24.9 &        13.0  &           -2.8 &       0.7 \\
                &  (4,  0) &                14.4 &  65 &     18.5 &        13.3  &           -4.7 &       0.8  & $\circ$\\
                &  (2,  0) &                15.4 &  51 &     22.4 &        14.2  &           -3.7 &       0.4 \\
    \hline
    \multicolumn{9}{l}{$^a$: intrinsic width of the line.}\\
    \multicolumn{9}{l}{$^b$: RV shift of the profile.}\\
    \multicolumn{9}{l}{$^c$: velocity amplitude of the pulsation.}\\
    \multicolumn{9}{l}{$^d$: $\bullet$ -- certain identification, $\circ$ -- ambiguous identification.}\\
  \end{tabular}
\end{table*}

\begin{table*}
  \centering
  \caption{Spectroscopic mode-identification results for fitting the Fourier-parameter profiles of the independent frequencies of the secondary component of \oo. See further column explanations in the footnotes of Table~\ref{tbl:spmodeidfullprim}.\label{tbl:spmodeidfullsec}}
  \begin{tabular}{lcrrrrrrc}
    \hline
    Method      & $(\ell, m)$ & $\chi^2_\mathrm{r}$ & $i$ & \vsini& $\sigma^{a}$ & $\Delta Z^{b}$ & $A(v)^{c}$ & adopted$^{d}$\\
                &             &                     &(deg)& \kms  & \kms         &    \kms        & \kms       \\
    \hline
    \multicolumn{9}{c}{$f^{\mathrm{S}}_1 = 0.020$\,\cd}\\[1.5mm]
    APf         &  (3, -2) &                73.5 &  10 &     18.8 &        17.4  &           33.1 &       2.8  & $\circ$\\[1.5mm]

    ZAPf        &  (4,  2) &                88.5 &  24 &     32.3 &        13.5  &           35.3 &       0.7 \\
                &  (3, -2) &                99.0 &  24 &     31.6 &        14.2  &           35.1 &       1.1  & $\circ$\\
                &  (4, -2) &               115.7 &  21 &     33.0 &        13.2  &           35.3 &       0.9 \\[1.5mm]

    APh         &  (2,  2) &                36.6 &  20 &     22.6 &        13.7  &           32.0 &       1.4 \\
                &  (3,  2) &                43.4 &  10 &     18.8 &        11.5  &           32.0 &       4.0 \\[1.5mm]

    ZAPh        &  (4,  2) &                46.7 &  26 &     32.3 &        12.5  &           36.1 &       0.6 \\
                &  (3, -2) &                53.7 &  19 &     33.0 &        12.5  &           35.3 &       1.6  & $\circ$\\
                &  (3,  2) &                55.4 &  26 &     33.0 &        12.1  &           35.9 &       0.8 \\
    \hline
    \multicolumn{9}{c}{$f^{\mathrm{S}}_2 = 0.159$\,\cd}\\[1.5mm]
    APf         &  (2, -2) &                21.5 &  14 &     19.1 &        17.1  &           32.8 &       2.9  & $\bullet$\\[1.5mm]

    ZAPf        &  (2, -2) &                46.2 &  26 &     33.0 &        13.6  &           35.2 &       1.1  & $\bullet$\\
                &  (2,  2) &                63.8 &  19 &     32.8 &        13.7  &           35.2 &       1.1 \\[1.5mm]

    APh         &  (2,  2) &                 4.4 &  12 &     18.0 &        17.7  &           33.1 &       0.7 \\
                &  (2, -2) &                 6.1 &  19 &     26.5 &        16.0  &           33.3 &       0.9  & $\bullet$\\[1.5mm]

    ZAPh        &  (2, -2) &                17.1 &  25 &     36.5 &        12.3  &           33.5 &       1.2  & $\bullet$\\
                &  (3,  2) &                19.2 &  15 &     32.4 &        12.7  &           36.0 &       1.9 \\
    \hline
    \multicolumn{9}{c}{$f^{\mathrm{S}}_3 = 0.232$\,\cd}\\[1.5mm]
    APf         &  (2, -2) &                16.5 &  16 &     20.0 &        13.9  &           33.5 &       3.7  & $\bullet$\\[1.5mm]

    ZAPf        &  (2, -2) &                43.8 &  27 &     33.0 &        13.9  &           35.0 &       1.2  & $\bullet$\\[1.5mm]

    APh         &  (2, -2) &                 3.6 &  15 &     20.8 &        13.4  &           32.8 &       4.8  & $\bullet$\\[1.5mm]

    ZAPh        &  (2, -2) &                24.8 &  31 &     32.8 &        12.5  &           36.6 &       0.8  & $\bullet$\\
                &  (3,  2) &                31.9 &  10 &     31.9 &        13.2  &           36.0 &       3.3 \\
                &  (4, -1) &                34.7 &   5 &     33.0 &        12.5  &           36.0 &       3.5 \\
    \hline
    \multicolumn{9}{c}{$f^{\mathrm{S}}_4 = 0.371$\,\cd}\\[1.5mm]
    APf         &  (2, -2) &                24.3 &  14 &     19.8 &        15.0  &           33.0 &       5.0  & $\bullet$\\
                &  (2,  2) &                33.8 &  14 &     18.1 &        17.5  &           33.3 &       0.2 \\[1.5mm]

    ZAPf        &  (2, -2) &                46.2 &  26 &     33.0 &        13.6  &           35.3 &       1.3  & $\bullet$\\
                &  (3,  2) &                61.0 &   9 &     32.9 &        13.4  &           35.3 &       3.8 \\
                &  (2,  2) &                62.5 &  10 &     33.0 &        13.5  &           35.3 &       3.3 \\[1.5mm]

    APh         &  (3,  2) &                 5.1 &   8 &     22.7 &        14.4  &           33.9 &       6.0 \\[1.5mm]

    ZAPh        &  (3,  2) &                14.5 &   9 &     33.0 &        12.5  &           35.9 &       4.2 \\
    \hline
    \multicolumn{9}{c}{$f^{\mathrm{S}}_5 = 1.191$\,\cd}\\[1.5mm]
    APf         &  (4,  1) &                21.5 &  24 &     18.9 &        14.8  &           31.6 &       0.8 \\
                &  (3,  2) &                26.1 &  17 &     29.7 &        12.2  &           28.0 &       5.1 \\[1.5mm]

    ZAPf        &  (3, -2) &                65.6 &  16 &     31.9 &        14.7  &           35.0 &       4.7  & $\circ$\\[1.5mm]

    APh         &  (3,  3) &                15.0 &  17 &     23.7 &        10.9  &           29.7 &       6.0 \\
                &  (2,  2) &                18.1 &  17 &     30.0 &         9.0  &           38.8 &       5.0 \\[1.5mm]

    ZAPh        &  (3, -2) &                54.7 &  21 &     31.3 &        13.6  &           36.6 &       2.2  & $\circ$\\
                &  (2, -2) &                67.5 &  34 &     26.3 &        15.0  &           38.6 &       2.4 \\
    \hline
    \multicolumn{9}{c}{$f^{\mathrm{S}}_6 = 1.129$\,\cd}\\[1.5mm]
    APf         &  (3, -2) &                13.9 &  10 &     30.8 &        18.9  &           33.1 &       5.2  & $\circ$\\
                &  (2,  2) &                16.9 &  17 &     33.0 &        12.7  &           32.2 &       6.2 \\
                &  (3,  2) &                20.8 &  19 &     30.8 &        14.2  &           31.8 &       6.0 \\[1.5mm]

    ZAPf        &  (3, -2) &                38.5 &  15 &     32.1 &        14.7  &           35.0 &       3.3  & $\circ$\\
                &  (1, -1) &                42.8 &  11 &     31.1 &        15.5  &           35.0 &       2.5 \\[1.5mm]

    APh         &  (3, -2) &                15.1 &  10 &     26.7 &        16.7  &           36.8 &       6.5  & $\circ$\\[1.5mm]

    ZAPh        &  (3, -2) &                23.1 &  12 &     33.0 &        13.4  &           35.4 &       4.3  & $\circ$\\
    \hline
  \end{tabular}
\end{table*}

\begin{table}
  \centering
  \caption{Inclination and rotation of the secondary component of \oo. Spectroscopic mode-identification results for simultaneously fitting the Fourier-parameter profiles of $f^{\mathrm{S}}_2, f^{\mathrm{S}}_3$ and $f^{\mathrm{S}}_3$ frequencies.\label{tbl:spmodeidsecondary}}
  \begin{tabular}{lcrrrrc}
    \hline
    Method      & $(\ell, m)$ & $\chi^2_\mathrm{r}$ & $i$ & \vsini & $\Delta Z^{*}$ & adopted$^{a}$\\
                &          &                     &(deg)& \multicolumn{2}{c}{(\kms)}\\
    \hline
    APf         &  (2, -2) &                24.2 &  17 &     23.6 & 33.1  & $\bullet$ \\[1.5mm]

    ZAPf        &  (2, -2) &                53.6 &  21 &     31.8 & 35.0  & $\bullet$ \\
                &  (2,  2) &                76.2 &   8 &     31.4 & 35.0 \\
                &  (3,  2) &                79.7 &   8 &     33.1 & 35.2 \\[1.5mm]

    APh         &  (2, -2) &                14.1 &  16 &     24.5 & 32.7  & $\bullet$ \\
                &  (3,  2) &                16.8 &  10 &     28.5 & 34.0 \\[1.5mm]

    ZAPh        &  (2, -2) &                26.8 &  20 &     34.6 & 34.8  & $\bullet$ \\
                &  (2,  2) &                32.6 &   8 &     37.6 & 32.3 \\
                &  (3,  2) &                38.4 &   9 &     36.8 & 32.9 \\
    \hline
    \multicolumn{6}{l}{$^{*}$: RV shift of the profile.}\\
  \end{tabular}
\end{table}

\subsubsection{Summary and discussion of the spectroscopic mode-identification results}
\label{sect:spmodeidsum}

\paragraph*{$\mathbf{f^{\mathrm{P}}_1: (1, -1)}$}
-- Successful identification. The spectroscopic result is supported by the results of the photometric identification of this mode. We used the ZAPh fitting of this frequency to determine the inclination and rotation properties of this component.

\paragraph*{$\mathbf{f^{\mathrm{P}}_2: (2, 2)\ \mathrm{or}\ (4, 2)}$}
-- Ambiguous identification. The results listed in Table~\ref{tbl:spmodeidfullprim} suggest that this frequency belongs either to a $(2, 2)$ or to a $(4, 2)$ mode. This frequency appears in the \most\/ light curve, but not in the Fairborn photometry, thus, photometric identification is not possible. A $(2, 2)$ mode is invisible from the direction of the equator and best visible from the poles, thus, its photometric detection is consistent with this solution. A $(4, 2)$ mode is invisible from 0\degr\ inclination, best visible from 40.9\degr, and invisible again from 67.8\degr. The detection of this frequency only in the space photometry but not in the ground-based data might also be consistent with this mode identification, as the inclination of this component is near the IACC, but no complete cancellation occurs.

\paragraph*{$\mathbf{f^{\mathrm{P}}_3: (4, 2)}$}
-- Successful identification. The ZAP profiles of this frequency are best fitted with a $(4, 2)$ mode by 3 of the 4 fitting methods, the only exception being APf. The photometric visibility discussed for $f^{\mathrm{P}}_2$ applies here as well, because this frequency is also detected only in the \most\/ light curve, but not in the Fairborn observations.

\paragraph*{$\mathbf{f^{\mathrm{P}}_4}$}
-- Unsuccessful identification. The spectroscopic mode-identification results are ambiguous. This periodicity is detected in the Fairborn light curves, and was identified as a probable $\ell=2$ mode. The spectroscopic identification provides only one such solution: a $(2, 0)$ mode is the second best with the APf fitting. However, the IACC of a $(2, 0)$ mode is 54.7\degr, almost equal to the inclination of the primary. Thus, even if this mode is an $\ell=2$ one, $m$ surely cannot be 0. Interestingly, the amplitude and phase profiles of this frequency and those of $f^{\mathrm{P}}_5$ are rather similar, suggesting that these might have the same $(\ell, m)$ modes. However, there is not much similarity between the best mode-identification solutions of these two.

\paragraph*{$\mathbf{f^{\mathrm{P}}_5: (4, 0)}$}
-- Ambiguous identification. Among the possible solutions, the $(4, 0)$ mode appears as the best fitting when the ZP profile is not fitted, and the second best when the ZP profile is also taken into account. However, this is the lowest-amplitude LPV of the primary, so there are only small differences in the goodness of the fit of the different modes. The photometric mode identification is also ambiguous for $f^{\mathrm{P}}_5$, also because of its low photometric amplitude. IACCs of a $(4, 0)$ mode are 30.6\degr and 70.1\degr, so a poorly visible variation of such a mode cannot be excluded at the inclination of the primary of $\approx$\,59\degr.

\paragraph*{$\mathbf{f^{\mathrm{S}}_1: (3, -2)}$}
-- Ambiguous identification. It is not even certain that this extremely-low-frequency variation originates from pulsation. However, due to the fast rotation of the secondary component, a retrograde, $m<0$ azimuthal-order mode might be able to explain the low pulsation frequency observed from the rest frame. If this is an $m=-2$ mode indeed, then its frequency in the co-rotating frame is $f^\mathrm{S}_\mathrm{1\,corot} = f^\mathrm{S}_1 - m f^\mathrm{S}_\mathrm{rot} = 1.6\pm 0.8$\,\cd, consistent with SPB pulsation. The APf fit yields the best fit, and both ZAP fits rank to the second place the $(3, -2)$ mode, which might be consistent with such a situation. Furthermore, all the other top results have positive $m$ values, which can definitely be excluded for the same reason. The overall high $\chi^2_\mathrm{r}$ values of the fits of the profiles of this frequency make the identification even more uncertain.

\paragraph*{$\mathbf{f^{\mathrm{S}}_2, f^{\mathrm{S}}_3\ \mathrm{and}\ f^{\mathrm{S}}_4: (2, -2)}$}
-- Successful identifications. The spectroscopic mode-identification results for $f^{\mathrm{S}}_3$ are the most univocal, but also for $f^{\mathrm{S}}_2$, only the APh method ranks this solution to the second place. For $f^{\mathrm{S}}_4$, $(2, -2)$ appears the best solution only with the full-profile fits, while fits of half of the profiles rank $(3, 2)$ as the best one, instead. However, taking into account the fast rotation of the secondary, the low frequency of this mode is not consistent with any $m>0$ azimuthal-order mode. The fact that these frequencies are not detected in photometry is in accordance with the identification of these modes, and with the obtained low inclination of the secondary component: sectoral ($m=\pm \ell$) modes have strong spatial cancellation when viewed from about the poles.

\paragraph*{$\mathbf{f^{\mathrm{S}}_5\ \mathrm{and}\ f^{\mathrm{S}}_6: (3, -2)}$}
-- Ambiguous identification. These two frequencies are discussed together, since their amplitude and phase profiles, as well as the fitting results, are quite similar. Oddly, the identification of the lower-amplitude $f^{\mathrm{S}}_6$ seems to be more certain. For this component, all four fitting methods of the spectroscopic mode identification yields the best fit with the $(3, -2)$ mode. This mode appears the best one also for the stronger pulsation of $f^{\mathrm{S}}_5$, however, only when the ZP is fitted. Based solely on the spectroscopic mode-identification results, we accepted the $(3, -2)$ solution for both frequencies. At the same time, we have to classify this result as ambiguous, since the photometric mode identification of both frequencies attributes the least probability for an $\ell=3$ mode. Also, the relatively high frequencies of these modes and the apparent fast rotation of the secondary are not quite compatible with these being $m=-2$ azimuthal-order retrograde modes.

\paragraph*{\bf Rotation and magnetic field}
-- Our spectropolarimetric analysis show the presence of a magnetic field in the secondary component of \oo, but not in the primary. At the same time, the analysis of the LPVs in the two components indicate fast rotation of the magnetic secondary component, and slow rotation of the non-magnetic primary. This is rather unexpected, because the magnetic field is assumed to slow down the rotation.

\paragraph*{\bf Period spacings of the three $(2, -2)$ modes of the secondary}
-- We also checked tentatively if the period spacings of the three modes of $f^{\mathrm{S}}_2$, $f^{\mathrm{S}}_3$ and $f^{\mathrm{S}}_4$ in the co-rotating frame agree with theoretical predictions of {\sc MAD} \citep{Dupret2001A&A...366..166D,Dupret2002A&A...385..563D}. Taking into account the estimated $\Omega = 0.8$\,\cd\ rotation frequency of the secondary (Sect.~\ref{sect:modeidsecondary}), the frequency shift caused by the rotation and the $m=-2$ retrograde propagation for an $\ell=2$ mode is about $\Omega m (1-1/(\ell(\ell+1))) = -1.3$\,\cd. Correcting with this shift, we obtain period spacings of 0.03 and 0.05\,d between the $(f^{\mathrm{S}}_2, f^{\mathrm{S}}_3)$ and $(f^{\mathrm{S}}_3, f^{\mathrm{S}}_4)$ modes, respectively. If a slightly lower rotation frequency of 0.6\,\cd, is considered, the correction is only $-1.0$\,\cd, and the period spacings are 0.05 and 0.08\,d. These latter values are more in accordance with the theoretically predicted period spacings for the stellar parameters obtained for the secondary component of \oo. The theoretical frequency spacings are around 0.045\,d in the range of the periods of these modes in the co-rotating frame (between 0.6 and 0.9\,d). In this case, $f^{\mathrm{S}}_2$ and $f^{\mathrm{S}}_3$ are consecutive radial-order modes, while the radial order difference between $f^{\mathrm{S}}_3$ and $f^{\mathrm{S}}_4$ is two. This result suggests that the rotation frequency of the secondary might be in the lower part of the uncertainty range given in Sect.~\ref{sect:modeidsecondary}, in accordance with the estimated small ($<10\degr$) correction effect on the inclination and equatorial rotation velocity due to the fast rotation, as discussed in the end of Sect.~\ref{sect:modeidsecondary}.

\section{Summary}

The results of our investigations have already been discussed in the previous sections, thus, here we only summarize our findings briefly.

\begin{itemize}
\item[$\bullet$]Spectroscopy shows that \oo\ is a double-lined binary star. The $O-C$ analysis of the dominant frequency shows that the orbital period is quite long: $\approx$\,8.9\,yr. No orbital solution could be derived from the present data.
\item[$\bullet$]The photometric $O-C$ analysis shows that the dominant frequency originates from the primary component.
\item[$\bullet$]Our investigation of the $O-C$ variations support the long-term phase coherence of the SPB pulsations over the whole investigated 20-yr time base.
\item[$\bullet$]We re-determined the mean physical parameters of the system from the published Geneva photometry of \oo\ \citep{decat07}, taking into account its metallicity of $\mathrm{[Fe/H]}=-0.3$\,dex \citep{Niemczura2003A&A...404..689N}: $T_\mathrm{eff} = 16\,600\pm800$\,K and $\log g = 4.22\pm0.2$\,dex.
\item[$\bullet$]Fitting the observed EW ratios of 21 individual metallic lines with synthetic spectra, and considering evolutionary tracks calculated with CL\'ES \citep{Scuflaire2008Ap&SS.316...83S}, we determined the following atmospheric parameters and masses of the two components of \oo: $T_\mathrm{eff}^\mathrm{P} = 16\,850\pm{800}\,\mathrm{K}, T_\mathrm{eff}^\mathrm{S} = 16\,250\pm{1000}\,\mathrm{K}, \log g^\mathrm{P} = 4.2\pm{0.2}, \log g^\mathrm{S} = 4.25\pm{0.25}, \log (L^\mathrm{P}/L_\odot) = 2.75\pm{0.29}, \log (L^\mathrm{S}/L_\odot) = 2.62\pm{0.36}, M^\mathrm{P} \approx 4.6\,M_\odot, M^\mathrm{S} \approx 4.2\,M_\odot$. Consequently, both components lie within the SPB instability region of the HRD.
\item[$\bullet$]Our spectropolarimetric observations indicate the presence of a magnetic field in the secondary component of \oo, while no magnetic signature was observed in the primary component. The polar field strength of the secondary is estimated to be of a few hundred Gauss.
\item[$\bullet$]We identified 11 independent significant frequencies and the second harmonic of the dominant frequency by Fourier analysis of different photometric and spectroscopic time-series. PbP Fourier analysis of the line profiles of the two components proved that, in accordance with their location in the HRD, both stars show LPVs consistent with stellar pulsations. With the PbP analysis, we were also able to relate each frequency to one of the binary components.
\item[$\bullet$]We performed photometric mode identification on the four-colour Str\"omgren light curves for the five frequencies identified in these data. Only the identification of the dominant mode is unambiguous. This one is most probably an $\ell=1$ mode. The other identifications are either poorly discriminative or are in contradiction with the spectroscopic mode identification.
\item[$\bullet$]Spectroscopic mode identification of the dominant frequency show that this belongs to an $(\ell, m) = (1, -1)$ mode. The mode identification of this frequency also yields the inclination and rotation parameters of this component: $i^\mathrm{P} = 59\degr\pm 5\degr$, $(v\sin i)^\mathrm{P} = 21.5 \pm 1.5$\,\kms. These show that the primary is a relatively slow rotator, since $v^\mathrm{P}_\mathrm{eq} = 25 \pm 2$\,\kms\ and $P^\mathrm{P}_\mathrm{rot} = 5.9 \pm 2$\,d.
\item[$\bullet$]Spectroscopic mode identification of $f^{\mathrm{S}}_2, f^{\mathrm{S}}_3$ and $f^{\mathrm{S}}_4$ give the inclination and rotation parameters of the secondary: $i^\mathrm{S} = 20\degr^{+7}_{-5}$, $(v\sin i)^\mathrm{S} = 35 \pm 4$\,\kms. These mean that the secondary component is a fast rotator: $v^\mathrm{S}_\mathrm{eq} = 100 \pm 30$\,\kms, $P^\mathrm{S}_\mathrm{rot} = 1.2 \pm 0.6$\,d. The fast rotation can explain the low observed frequencies of these three modes and especially the extremely low frequency of $f^{\mathrm{S}}_1 = 0.020$\,\cd.
\item[$\bullet$]The magnetic field measurements and the rotation speeds of the two components show just the opposite relation of what we would expect. The magnetic secondary rotates faster than the non-magnetic primary, while the magnetic field is assumed to slow down the rotation.
\item[$\bullet$]The rotation axes of the two components are probably misaligned by as much as 30$\degr$.
\item[$\bullet$]We performed spectroscopic mode identification for all the frequencies detected in the LPVs. The identification of 5, 5 and 1 modes were successful, ambiguous and unsuccessful, respectively.
\end{itemize}

Detailed theoretical asteroseismic modelling of the two components of the \oo\ system is planned to be the topic of another paper in the future.

\section*{Acknowledgements}

\'A.S. is grateful to Yves Fr\'emat at the Royal Observatory of Belgium for the tutorial and discussions about the calculation and usage of synthetic stellar spectra.
The authors are indebted to Frank Fekel and Matthew Muterspaugh at the Tennessee State University for their help in organizing the Fairborn observations.
We thank the anonymous referee for their remarks, which helped improving this paper.

\'A.S. acknowledges support of the Belgian Federal Science Policy (project M0/33/029, PI: P.D.C.).
\'A.S. and Zs. B. acknowledges support of the E\"otv\"os Scholarship from the Hungarian Scholarship Board Office.
M.B. is F.R.S.-FNRS Postdoctoral Researcher, Belgium.
G.W.H. and  M.H.W. acknowledge support from NASA, NSF, Tennessee State University, and the State of Tennessee through its Centers of Excellence program.
T.K. acknowledges financial support from the Austrian Science Fund (FWF P23608).
P.G.B. was supported through the European Research Council under the European Community's Seventh Framework Programme (FP7/2007--2013)/ERC grant agreement n$^\circ$227224 (PROSPERITY).
C.A.E. and R.J.C. acknowledge financial support from NRF and University of Johannesburg.
J.N.F. acknowledges the support from the Joint Fund of Astronomy of National Natural Science Foundation of China (NSFC) and Chinese Academy of Sciences through the key project Grant U1231202, and the partial support from the National Basic Research Program of China (973 Program 2013CB834900).
S.Sc. acknowledges funding from the FWO Pegasus Marie Curie Fellowship program.
J.M. acknowledges a PhD fellowship of the Research Foundation -- Flanders (FWO).

We used the software package \famias\ for this study, which was developed in the framework of the FP6 European Coordination Action HELAS\footnote{\tt http://www.helas-eu.org/}.

Based on observations made with
the Mercator telescope (HERMES spectrograph) at the Roque de los Muchachos observatory (La Palma, Canary Islands),
the 2-m AST (T13), 0.75-m APT (T5) and 0.4-m APT (T3) at Fairborn Observatory (Arizona, USA),
the McLellan telescope (HERCULES spectrograph) at the Mount John University Observatory (New Zealand),
the Otto Struve telescope (SES Spectrograph) at the McDonald Observatory (Texas, USA),
the 1.9-m telescope (GIRAFFE spectrograph) and 0.5-m telescope (MP photometer) at the South African Astronomical Observatory (South Africa),
the 2.16-m telescope (Coude spectrograph) at the Xinglong Observatory (China),
the 1.5-m telescope (GAOES spectrograph) at the Gunma Astronomical Observatory (Japan),
the 2-m Alfred Jensch telescope (Coude echelle spectrograph) at the Th\"uringer Landessternwarte Tautenburg (Germany),
the 1.9-m telescope (HIDES spectrograph) at the Okayama Astrophysical Observatory (Japan),
the 1.2-m telescope (McKellar spectrograph) at the Dominion Astronomical Observatory (Canada),
the 1.5-m telescope (AURELIE spectrograph) at the Observatoire Haut Provance (France),
the Bernard Lyot telescope (NARVAL spectropolarimeter) at the Observatoire du Pic du Midi (France),
the Canadian-France-Hawaii Telescope (ESPaDOnS spectropolarimeter) at the Mauna Kea observatory (Hawaii),
the Euler telescope (CORALIE spectrograph) at the European Southern Observatory (La Silla, Chile),
and the Canadian MOST satellite.

\bibliography{sodor_hd25558_rev.bib}{}

\label{lastpage}

\end{document}